\documentclass[aps,prb, reprint,superscriptaddress]{revtex4-2}

\renewcommand{\selectlanguage}[1]{}

\usepackage{amsmath}
\usepackage[utf8]{inputenc}
\usepackage{physics}
\usepackage{enumitem}
\usepackage{float}
\usepackage[margin = 2cm]{geometry}
\usepackage{color}
\usepackage{comment}
\usepackage{graphicx}
\usepackage{dsfont}
\usepackage{soul}

\usepackage{appendix}
\usepackage{braket}
\usepackage{chemformula}
\usepackage{dcolumn}
\usepackage{bm}
\usepackage{setspace}
\usepackage{xcolor}

\usepackage{tikz}

\usepackage{braket}
\usepackage{graphicx}
\usepackage{dcolumn}
\usepackage{bm}
\usepackage[colorlinks=true, citecolor=blue, linkcolor=blue, urlcolor=blue]{hyperref}
\usepackage{indentfirst}

\definecolor{blue1}{HTML}{3a5377}
\definecolor{purple1}{HTML}{83606f}
\definecolor{red1}{HTML}{db705f}
\definecolor{black1}{HTML}{021225}

\definecolor{purple2}{HTML}{5d5d78}
\definecolor{red2}{HTML}{c2655f}
\definecolor{yellow1}{HTML}{e5bd91}

\definecolor{purple3}{HTML}{88606d}
\definecolor{red3}{HTML}{e98c6a}

\definecolor{purple4}{HTML}{818BA1}
\definecolor{purple5}{HTML}{A88F9A}

\DeclareRobustCommand\sampleline[1]{%
  \tikz\draw[#1, line width=1.8pt] (0,0) (0,\the\dimexpr\fontdimen22\textfont2\relax)
  -- (1.9em,\the\dimexpr\fontdimen22\textfont2\relax);%
}

\DeclareRobustCommand\dashedline[1]{%
  \tikz\draw[#1, line width=1.8pt] (0,0) (0,\the\dimexpr\fontdimen22\textfont2\relax)
  -- (1.9em,\the\dimexpr\fontdimen22\textfont2\relax);%
}

\def \ETH{Institute for Theoretical Physics, ETH Z\"urich, CH-8093 Z\"urich, Switzerland}
\begin{document}

\title{Bipolaron dynamics in the one-dimensional SSH model}

\author{Filip Marijanovi\'c}
\affiliation{\ETH}

\author{Yi-Fan Qu}
\affiliation{\ETH}

\author{Eugene Demler}
\affiliation{\ETH}

\date{\today}

\begin{abstract}

Characterizing bipolaron binding, and understanding how it depends on electron-phonon interaction, is crucial to unraveling the nature of emergent many-body states in strongly interacting electron-phonon systems. So far, most studies of bipolarons have been limited to the Holstein model, in which the coupling constant is momentum-independent. The paradigmatic example of momentum-dependent electron-phonon interaction comes from the system in which phonon distortions modify electron hopping, the SSH model. Already individual polarons in the SSH model are richer than the Holstein model counterparts, and feature a phase transition into the finite momentum ground state with increasing electron-phonon interaction. In this paper, we use a variational approach to study bipolarons in the one-dimensional SSH model and discuss their ground state, dispersion, and excitation spectra. We explore the full parameter range of the system, including the adiabatic regime of slow phonons, which was inaccessible to previous theoretical studies. In agreement with earlier studies, we find that in the anti-adiabatic strongly interacting regime, bipolarons have low effective mass. By contrast, in the adiabatic case, we find that increasing electron-phonon interactions results in an exponential increase of the bipolaron mass. We establish the existence of multiple branches of bound excited states of SSH bipolaron and discuss the signatures of these bound states in dynamics. We show that in the anti-adiabatic regime, response functions obey a parity selection rule, that imposes symmetry constraints on the excitation spectra and provides a clear signature of SSH bipolarons.
\end{abstract}

\maketitle

\section*{Introduction}

Electron-phonon interactions are ubiquitous across the entire field of condensed matter physics ~\cite{franchini_polarons_2021} and give rise to paradigmatic phenomena such as superconductivity (SC)~\cite{PhysRev.108.1175}, charge density waves (CDW)~\cite{10.1093/acprof:oso/9780198507819.001.0001,RevModPhys.60.1129,bonca_mobile_2000}, and ferroelectricity~\cite{Rabe2007PhysicsOF,PhysRevB.35.4840}.  
Historically, analysis of interacting electron-phonon systems assumed momentum independent coupling constant, such as in the Holstein model ~\cite{hohenadler_quantum_2004,alexandrov_polaronic_1992,austin_polarons_1969,PhysRevB.45.13109,10.1143/PTP.26.29,chakraborty_stability_2012,alexandrov_advances_2010, alexandrov_polaron_2000, devreese_frohlich_2009, devreese_frohlich_2015,kuroda_dynamics_1985,aydin2024polaronformationquantumacoustics,PhysRevB.106.054311, thomas2024theoryelectronphononinteractionsextended}.  However, recent theoretical and experimental studies suggest that in many strongly correlated materials, such as high $T_c$ superconductors ~\cite{mierzejewski_eliashberg_1996,rosch_electron-phonon_2004,devereaux_neutron_1999,song_electron-phonon_1995,heid_momentum_2008,bulut_d_x2ensuremath-y2_1996,johnston_systematic_2010,devereaux_anisotropic_2004,altendorf_temperature_1993-1,friedl_determination_1990,devereaux_directly_2016,cuk_review_2005} as well as in certain CDW materials~\cite{doran_calculation_1978,eiter_alternative_2013,kurzhals_electron-momentum_2022,pouget_momentum-dependent_2021}, the electron-phonon interaction has strong momentum dependence and is anisotropic near the Fermi surface. Theoretical analysis also suggests that anisotropic electron-phonon interactions are more robust against screening by Coulomb interactions, leading to potentially larger couplings and qualitatively different physics~\cite{johnston_systematic_2010,devereaux_anisotropic_2004}. For example, they are expected to favor unconventional electron pairing, such as d-wave superconductivity in high Tc cuprates ~\cite{johnston_systematic_2010, song_electron-phonon_1995,marchand_sharp_2010}. 

Differences in the effects of Holstein~\cite{adamowski_formation_1989,hai_electron_2018,chandler_extended_2014,robinson_cumulant_2022} and SSH phonons on electrons can be appreciated already in the weakly interacting, perturbative regime. In this limit, phonons can be integrated out, resulting in an effective phonon-mediated interaction between electrons.  SSH phonons couple to electron hopping rather than on-site energies ~\cite{barisic_rigid-atom_1972,heeger_solitons_1988,su_solitons_1979, hicks_lattice_1985}, hence, they give rise to the effective interaction that is spatially more non-local and includes correlated pair hopping ~\cite{PhysRevB.89.144508, marchand_sharp_2010,sous_light_2018}. The difference between the two types of electro-phonon couplings becomes even more pronounced in the strongly interacting regime when each electron gives rise to a large local occupation of phonons. Strong polaronic dressing modifies dispersion of both polarons and bipolarons~\cite{bassani_variational_1991,perroni_polaron_2004,chakraverty_experimental_1998,senger_stability_nodate,sil_formation_1999,hai_electron_2018}, changes binding energies of bipolarons, and ultimately determines competition between the localized and extended states of bipolarons. The latter determines whether many-electron states have dominant superconducting or CDW instabilities, or even exhibit phase separation ~\cite{PhysRevB.40.197,PhysRevB.99.174516,PhysRevLett.120.187003}. It is natural to expect that changing the character of electron-phonon coupling results not only in quantitative change of the SC and CDW transition temperatures but also in the nature of many-body states. So far, Holstein-type models in the strongly interacting regime have received much more attention theoretically, including detailed analysis of bipolarons ~\cite{devreese_frohlich_2009,alexandrov_polaron_2000,lin_analysis_2021,sio_polarons_2019}. The quest to understand bipolarons in the strongly interacting regime of the SSH model in both adiabatic (slow phonon) and anti-adiabatic (fast phonon) regimes is the primary motivation of this paper.

Recent theoretical studies of the electron-phonon systems with momentum dependent couplings analyzed the Bond-Peierls ~\cite{carbone_bond-peierls_2021,zhang_bipolaronic_2023,zhang_bond_2022,malkaruge_costa_comparative_2023} and SSH~\cite{PhysRevB.89.144508,marchand_sharp_2010, shi_variational_2018,zhang_peierlssu-schrieffer-heeger_2021,Banerjee_2023,banerjee2024identifyingquantifyingsuschriefferheegerlikeinteractions} type models in one-dimensional systems. In both cases, phonon displacements couple to electron hopping and the two only differ in whether phonon modes are located on lattice sites or at midpoints of the bonds. These studies revealed that individual polarons in both models are significantly lighter than those in the Holstein model. Ground state properties of bipolarons have been studied in the SSH model but only in the anti-adiabatic regime, when phonon frequency is higher than the single electron hopping. In this regime, high phonon frequency suppresses large occupation numbers of phonons and makes theoretical analysis more tractable. Ref~\cite{sous_light_2018} found that in this case bipolarons also remain light. Properties of the SSH bipolarons in the adiabatic regime, including their effective mass, as well as the dynamical properties of the SSH bipolarons, have not been previously explored. 

\begin{figure}[H]
    \centering
    \includegraphics[width=1.0\linewidth]{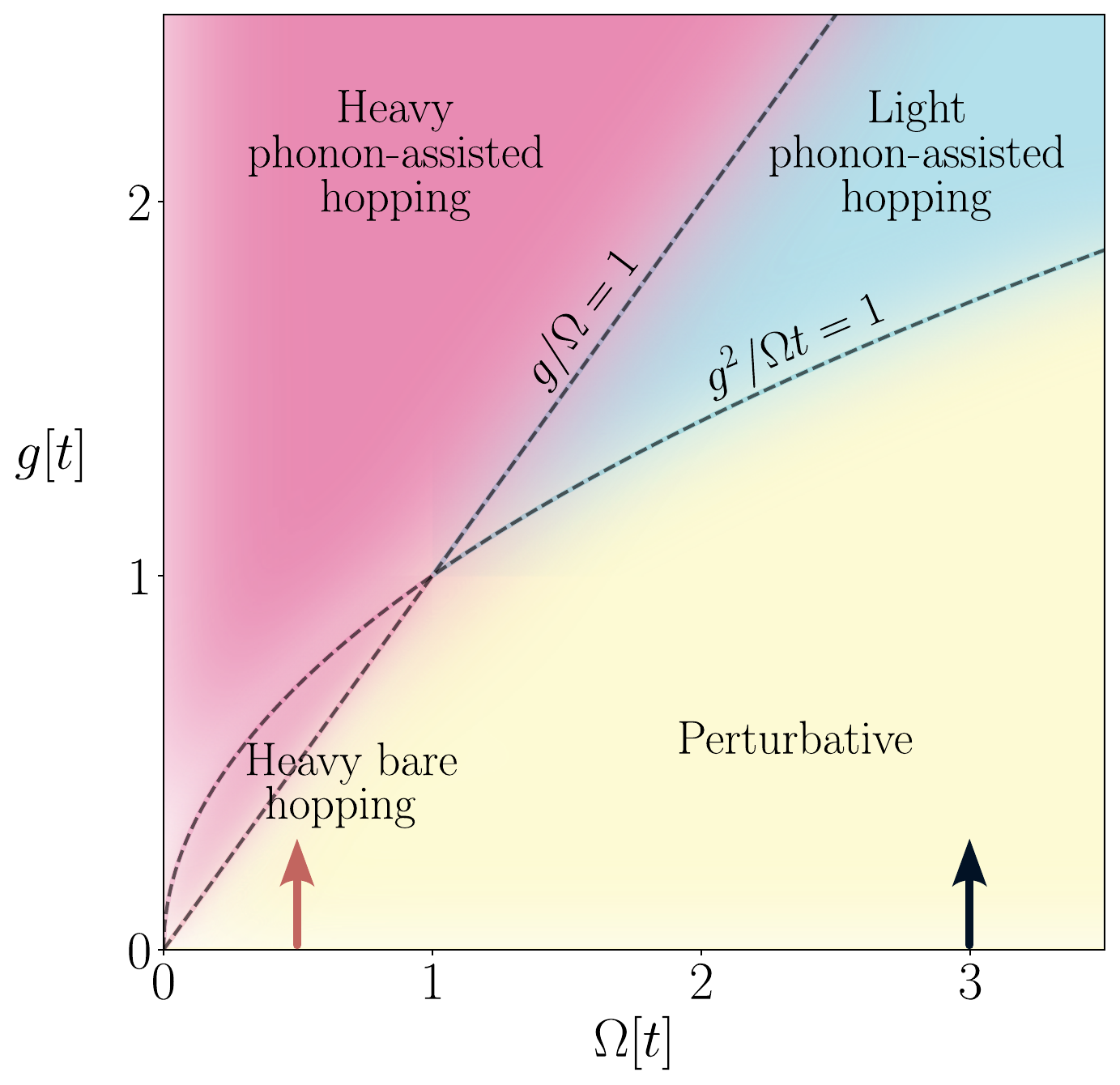}
    \caption{Qualitative diagram of the different bipolaron regimes within the SSH model as a function of the phonon frequency $\Omega$ and electron-phonon coupling $g$, in units of electron hopping $t$. The dashed lines correspond to boundaries between different regimes $g^2/\Omega t = 1 $ and $g/\Omega = 1$. The arrows indicate the vertical cuts explored in this work, the adiabatic $\Omega = 0.5t$ and the anti-adiabatic $\Omega = 3.0t$ regimes. Different regimes indicate whether the hopping is dominated by phonon-assisted or bare processes and whether the effective mass is large (heavy bipolaron) or small (light bipolaron). In the heavy regimes, the effective mass grows exponentially with phonon numbers i.e. $g/\Omega$.}
    \label{fig:diagram}
\end{figure}

\begin{figure}[H]
    \centering
    \includegraphics[width=1.0\linewidth]{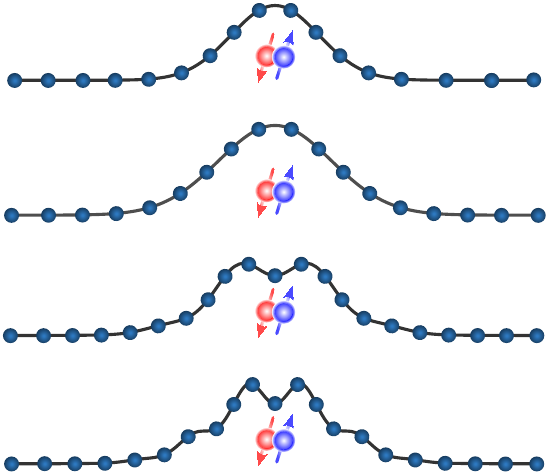}
    \caption{Sketch of the bound excited states within the SSH model at strong coupling. The excitations correspond to breathing/ripple deformation of the phonon cloud around the bipolaron, while the relative electron coordinate does not change considerably. This is in direct contrast with the usual excitations in the Holstein model.}
    \label{fig:excitations}
\end{figure}

This paper provides a comprehensive study of bipolarons in a one-dimensional SSH model. We develop a variational ansatz using the Lee-Low-Pines transformation~\cite{lee_motion_1953}, which allows us to study the ground state and the excited state properties, as well as the low-energy response and spectral functions. We use a superposition of coherent states of phonons in the LLP frame as a basis and explore the full parameter regime of the model, including the adiabatic regime with large phonon numbers. This regime was previously difficult to access~\cite{sous_light_2018} but is relevant for the majority of experimentally realizable systems. Due to the computational efficiency of the coherent states, our method can also easily be generalized to two-dimensional systems, going beyond the usual limitations of previous methods~\cite{sous_light_2018, marchand_sharp_2010}.  This paper presents three main results. Firstly, we demonstrate that the dependence of the effective mass of bipolarons on the electron-phonon coupling constant is very different in the adiabatic and anti-adiabatic regimes. Whereas, in the anti-adiabatic regime the bipolaron remains light, in the adiabatic regime the effective mass increases exponentially at strong coupling. 
A qualitative diagram, describing the behavior of the effective mass is shown in Figure \ref{fig:diagram}, as a function of the electron-phonon coupling $g$ and the phonon frequency $\Omega$, in units of electron hopping $t$. Different parameter regimes can be distinguished by two dimensionless parameters; (i)  $g^2/ \Omega t$, describing the relative contribution of the bare and phonon-assisted hopping processes; (ii) $g/\Omega$, describing the phonon occupation numbers. The importance of the former parameter can be understood by considering the two different hopping processes, bare and phonon-assisted. Since the phonon number is given by $g/\Omega$, the relevant energy scale for phonon-assisted hopping is $g^2/\Omega$, while for bare hopping it is $t$, making the ratio $g^2/\Omega t $. At large phonon numbers, the effective mass of the bipolaron grows exponentially, as in the Holstein model. Since the bipolaron hopping involves moving the phonon cloud, the hopping gets suppressed exponentially in the phonon number and hence the dimensionless coupling $g/\Omega$. Therefore, the phase diagram can be split into four different regions, depending on whether the hopping is predominantly phonon-assisted or not, and whether the phonon occupation is large or not. The striking difference between the adiabatic and the anti-adiabatic regimes can now be simply understood. The adiabaticity parameter $\Omega/t$ determines which vertical cut along the diagram the bipolaron will experience. In the adiabatic regime, increasing the coupling leads to an exponential growth of the effective mass immediately upon exiting the pertubative regime, while in the anti-adiabatic regime, there is a light, phonon-hopping dominated regime and the exponentially heavy regime is delayed. Note, that the quantitative nature of the diagram and the sharpness of certain transitions depends on the microscopic details of the model, but the qualitative features remain the same as long as the phonons couple linearly to the electron hopping. 
Secondly, we present the excitation spectrum, which consists of multiple discrete excited bands, a single-phonon continuum, a mixed particle-hole and phonon continuum, and an unbound polaron continuum. The discrete bands have energies below a single-phonon continuum and correspond to a breathing/ripple excitation of the phonon cloud on top of the bipolaron, while the relative electron coordinate does not change significantly(see Figure \ref{fig:excitations}). These modes share similarities with the \textit{relaxed excited states} of the Holstein model\cite{devreese_frohlich_2015}, where the electronic orbital state and the phonon cloud both deform self-consistently. However, in the SSH model, the discrete states have energies below the single-phonon continuum and involve a less dramatic change of the electronic orbitals, corresponding to mostly phononic excitations. 
Finally, we compute the response of the bipolarons to an external drive coupling to the electron density, which corresponds to an optical excitation of the system. We find that in the anti-adiabatic regime, there is an effective \textit{parity selection rule}, which arises as a consequence of the nearest-neighbor hopping nature of the SSH electron-phonon coupling. Along with the excitations below a single phonon continuum, the \textit{parity selection rule} provides a clear signature of the bipolaron formation in the SSH-like momentum-dependent electron-phonon coupling.

The paper is organized as follows. We begin by introducing the SSH model in Section \ref{sec:model}, and utilize symmetries of the model to construct an efficient variational ansatz. In Section \ref{sec:GS}, we discuss the ground state properties of the bipolaron and in Section \ref{sec:EXC}, we discuss the properties of excited states. Finally, in Section \ref{sec:response}, we construct the linear response functions, providing evidence of bipolaron formation. We conclude in Section \ref{sec:conclusion}, discussing the main results and possible future directions.

\section{Model and methods}
\label{sec:model}
We consider a one-dimensional lattice model with two electrons of opposite spins with tight binding dispersion,  and the SSH type of electron-phonon interaction.  Phonon modes are defined on lattice sites and couple to the tunneling of electrons between the corresponding sites and their nearest neighbors.  Hamiltonian of the system can be written as

\begin{equation}
    \hat{H} = \hat{H}_{\rm e} + \hat{H}_{\rm ph} + \hat{H}_{\rm e-ph},
    \label{eqn:Hamiltonian}
\end{equation}
where $\hat{H}_{\rm e}$ describes the nearest neighbor hopping of free electrons with amplitude $t$
\begin{equation}
    \hat{H}_{\rm e} = -t\sum_{n,\sigma} (\hat{c}^\dag _{n,\sigma} \hat{c}_{n+1,\sigma} + \hat{c}^\dag_{n+1,\sigma} \hat{c}_{n,\sigma}),
\end{equation}
and $\hat{c}_{n,\sigma}^\dag(\hat{c}_{n,\sigma})$ is the electron creation (annihilation) operator at the lattice site $n$ with spin $\sigma$. $\hat{H}_{\rm ph} = \Omega \sum_n \hat{b}_n^\dag \hat{b}_n $ is the phonon Hamiltonian, describing optical/Einstein phonons at frequency $\Omega$ and  $\hat{b}_n^\dag(\hat{b}_n)$ is the phonon creation (annihilation) operator at the lattice site $n$. The electron-phonon coupling is given by

\begin{equation}
    \hat{H}_{\rm e-ph} =g \sum_{n,\sigma} (\hat{c}^\dag _{n,\sigma} \hat{c}_{n+1,\sigma} + \hat{c}^\dag_{n+1,\sigma} \hat{c}_{n,\sigma} )(\hat{x}_{n+1} - \hat{x}_{n}),
    \label{eqn:SSH-coupling}
\end{equation}
where $g$ is the electron-phonon coupling strength and $\hat{x}_n = \hat{b}_n + \hat{b}^\dag_n$ is the lattice displacement. This model has been originally introduced by Su, Schrieffer, and Hegger in the context of polyacetylene~\cite{heeger_solitons_1988,su_solitons_1979}, but subsequent studies suggested that it can be used to describe other types of electron-phonon systems as well~\cite{barisic_rigid-atom_1972}. The physical origin of the model can be understood by considering the construction of the tight-binding Hamiltonian. Generally, the electron hopping amplitude depends on the orbital overlap of neighboring Wannier states. Therefore, changing the distance between the ions by displacing the lattice changes the Wannier orbital overlap and modifies the hopping amplitude, which is given by \eqref{eqn:SSH-coupling}, up to linear order ~\cite{barisic_rigid-atom_1972,heeger_solitons_1988}.

In momentum space the SSH interaction can be re-written as $\sum_{k,q,\sigma} g({k,q}) \hat{c}^\dag_{k+q,\sigma} \hat{c}_{k,\sigma} \hat{b}_q + \text{H.c.} $, where $g(k,q) = 2ig (\sin(k) - \sin(k+q))$,  $\hat{c}^\dag_{k,\sigma}(\hat{c}_{k,\sigma})$ are the electron creation(annihilation) operators and $\hat{b}^\dag_k(\hat{b}_k)$ are the phonon creation(annihilation) operators at momentum $k$, and we take the lattice constant to be unity. The SSH model provides a simple example of momentum-dependent electron-phonon coupling. This should be contrasted with the  Holstein model $\hat{H}_{\rm hol} = g_{h} \sum_{n,\sigma} \hat{c}^\dag_{n,\sigma} \hat{c}_{n,\sigma} \hat{x}_n = \sum_{q,\sigma} g(q) \hat{c}^\dag_{k+q,\sigma} \hat{c}_{k,\sigma} b_q + h.c. $, where $g(q) = g_h$ and its long-range generalizations ~\cite{gerlach_analytical_1991,alexandrov_advances_2010,alexandrov_polaronic_1992,austin_polarons_1969,chakraborty_stability_2012, goodvin_momentum_2008}, where the interactions couple to the local electron density.  In recent years, another general model with momentum-dependent coupling has been considered, the Bond-Peierls model  ~\cite{carbone_bond-peierls_2021,zhang_bipolaronic_2023,zhang_bond_2022,malkaruge_costa_comparative_2023}. The main difference to the SSH model is that the phonons are placed on lattice bonds, which changes the form of $g(k,q)$. As a result, the single polaron properties are very different compared to the SSH model ~\cite{marchand_sharp_2010,shi_variational_2018,malkaruge_costa_comparative_2023} and the exact strong coupling solution can be obtained by integrating out the phonon degrees of freedom ~\cite{carbone_bond-peierls_2021}. Unlike the Holstein and the Peierls bond model, in the SSH model, not even the single polaron strong coupling solution is known, making the numerical approaches very challenging.

\subsection{Hamiltonian Symmetries}

To understand the properties of the SSH model, we first review its symmetries. When we impose the periodic boundary conditions, the whole system is translationally invariant, so the total momentum, including both the electrons and phonons, is conserved modulo $2\pi$. The lattice structure also ensures that all properties are $2\pi$ periodic and the inversion symmetry implies that the observables do not depend on the sign of total momentum. The system also has a time-reversal symmetry, which conjugates any imaginary parts and leaves the positions invariant. The exact form is $K: i \rightarrow - i, \hat{c}_{n,\sigma} \rightarrow \Bar{\sigma} \hat{c}_{n,\Bar{\sigma}}$, where $\Bar{\sigma}$ is the spin opposite to $\sigma$ and we take $\sigma = \pm 1$ for spin up/down respectively. Phonon operator $\hat{b}_n$ is left invariant under conjugation. Finally, the system is symmetric under particle exchange $\hat{c}_{n,\uparrow} \leftrightarrow e^{i\theta} \hat{c}_{n,\downarrow}$, with $\theta$ being any phase. This follows directly from the full $SU(2)$ symmetry of the Hamiltonian, which has no explicit spin dependence. 

\begin{figure}[H]
    \centering
    \includegraphics[width=1.0\linewidth]{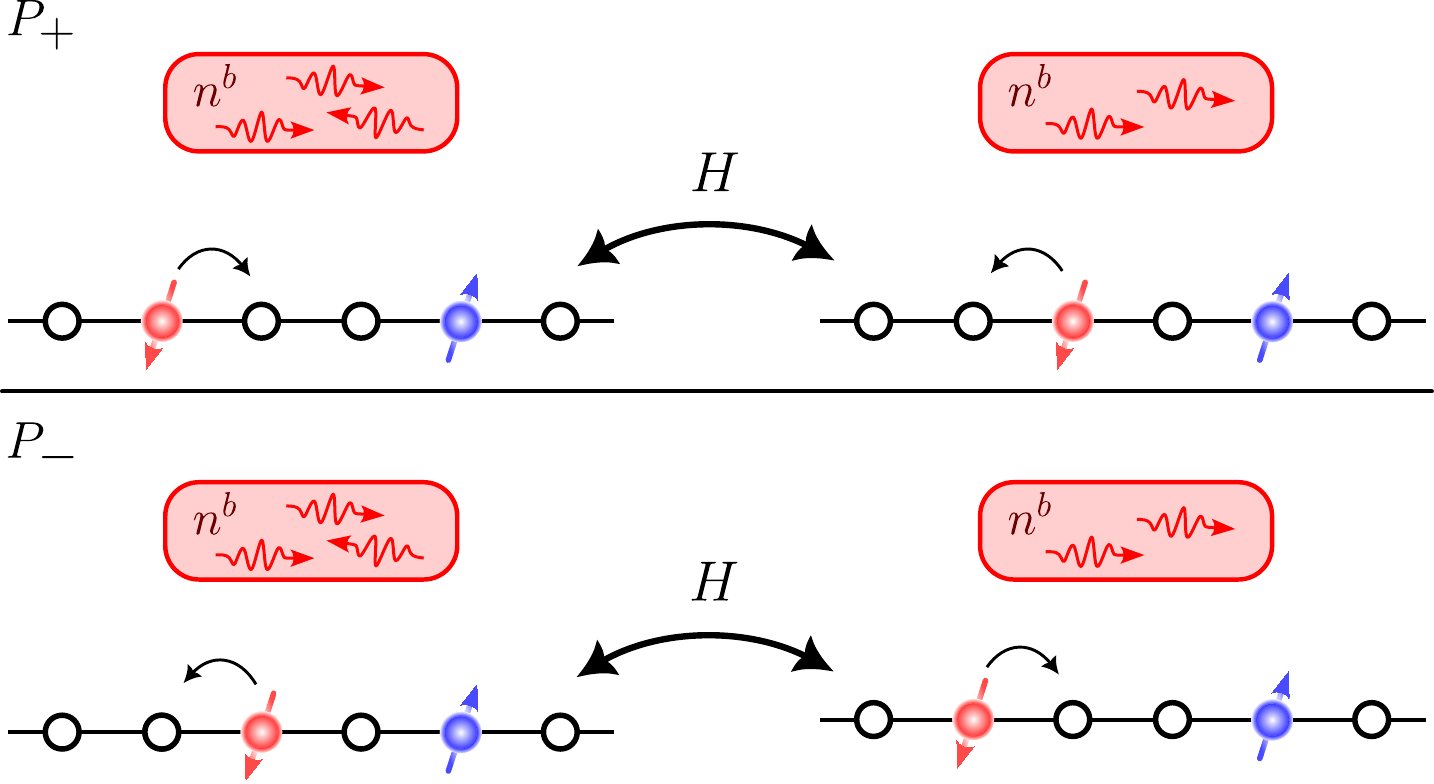}
    \caption{Two parity sub-spaces $P_+$ and $P_-$ of the full Hilbert space, which are disconnected in the zero hopping limit $t=0$. a) In $P_+$ sub-space, only two types of states are allowed: (left) odd electron distance with an odd phonon number $N_{ph}$, (right) even electron distance with an even phonon number. b) In $P_-$ sub-space, only two types of states are allowed: (left) even electron distance with an odd phonon number $N_{ph}$, (right) odd electron distance with an even phonon number.}
    \label{fig:parity-sketch}
\end{figure}

Apart from the symmetries of the full Hamiltonian, there is also an additional symmetry in the zero bare hopping limit $t = 0$. Since the electrons can only hop by emitting or absorbing a phonon, there is an effective \textit{parity} symmetry (See Figure \ref{fig:parity-sketch}). The electron relative position plus the total phonon number are conserved modulo 2.   In the operator form, the parity operator can be written as $\hat{P} = (-1)^{\hat{X}_{\uparrow} - \hat{X}_{\downarrow} + \hat{n}^b}$, where $\hat{n}^b = \sum_q \hat{b}^\dag_q \hat{b}_q$ is the total phonon number operator and $\hat{X}_\sigma = \sum_n n \hat{c}_{n,\sigma}^\dag \hat{c}_{n,\sigma}$ is the electron position operator. Intuitively, if electrons start an odd distance apart with an even number of phonons in the system, to change the relative distance by 1 they have to change the total phonon number by 1, therefore changing the relative position plus the total phonon number by 2. As a result, the Hilbert space can be split into two distinct parity sectors $P_+$ with total distance plus phonon number being even and $P_-$ with total distance plus phonon number being odd. For finite hopping, the exact parity symmetry is broken, however, in the strong coupling regime, it becomes important, as the effect of bare hopping gets suppressed. 

\subsection{Unitary transformations of the Hamiltonian}

Unitary transformations are commonly used in theoretical analysis of many-body quantum systems to simplify the Hamiltonian and utilize symmetries of the problem ~\cite{Wagner1986}. The model we consider has translational invariance, which makes it possible to use the Lee-Low-Pines (LLP) transformation ~\cite{lee_motion_1953} to separate the total system momentum $\kappa$ as a constant of motion. The two-particle LLP transformation can be done as follows
\begin{equation}
  \hat{U}_{\rm LLP} = e^{-i \hat{X}_\uparrow (\hat{P}_\downarrow + \hat{Q}_b)}, 
  \label{eqn:LLP}
\end{equation}
where $\hat{X}_\uparrow = \sum_n n \hat{c}_{n,\uparrow}^\dag \hat{c}_{n,\uparrow}$, $\hat{P}_\downarrow = \sum_k k \hat{c}^\dag_{k,\downarrow} \hat{c}_{k,\downarrow}$ and $\hat{Q}_b = \sum_q q \hat{b}_q^\dag \hat{b}_q$. The transformation shifts the coordinate system into the co-moving frame of the spin-up electron, where its momentum is conserved, with the resulting Hamiltonian $\sum_\kappa \hat{H}_{\rm{LLP}\uparrow}(\kappa) \hat{c}^\dag_{\kappa,\uparrow} \hat{c}_{\kappa,\uparrow}$ (see Appendix \ref{sec:Hamiltonian}). This reduces the problem to a single electron subspace, allowing it to be solved for each value of total momentum $\kappa$ independently. The LLP transformation \eqref{eqn:LLP} explicitly breaks the particle exchange symmetry. To restore this symmetry, we perform an additional unitary transformation 
\begin{equation}
    \hat{U}_{\rm LLP:2} = e^{i \pi \sum_q \hat{b}^\dag _q \hat{b}_q /2} e^{-\frac{i}{2} \hat{X}_\downarrow (\hat{Q}_b - \kappa)},
\end{equation}
where  $\hat{X}_\downarrow = \sum_n n \hat{c}_{n,\downarrow}^\dag \hat{c}_{n,\downarrow}$ . The exponential $e^{-\frac{i}{2} \hat{X}_\downarrow \hat{Q}_b}$ shifts the coordinate system halfway between the two electrons, and the other exponential factors are used for later convenience. The final Hamiltonian then becomes

\begin{widetext}
   \begin{equation}
    \begin{split}
        \hat{H}(\kappa) &= \Omega \sum_q \hat{b}^\dag _q \hat{b}_q - \sum_{n,\delta = \pm 1} 2 t \cos\left( \frac{\hat{Q}_b - \kappa}{2}\right) \hat{c}^\dag _{n-\delta,\downarrow} \hat{c}_{n,\downarrow}          -\frac{4g}{\sqrt{N}} \sum_{q,n,\delta} \sin(q/2) \hat{c}^\dag_{n-\delta,\downarrow} \hat{c}_{n,\downarrow} \times \\ &\left[ \cos\left(\frac{\hat{Q}_b - \kappa}{2}\right)\cos\left( \frac{q}{2}(1-n\delta)\right) (\hat{b}_q + \hat{b}^\dag_{q}) \right.
        - \left.\sin\left( \frac{\hat{Q}_b - \kappa}{2}\right) \sin\left( \frac{q}{2} (1-n\delta)\right) (\hat{b}_q -\hat{b}_{q}^\dag) \right],
    \label{eqn:Hamiltonian_final}
    \end{split}
\end{equation} 
\end{widetext}
which can be related to the initial Hamiltonian as $\hat{H} = \hat{U}_{\rm LLP} \hat{U}_{\rm{LLP:2}} \left( \sum_\kappa  \hat{H}(\kappa) \hat{c}^\dag_{\kappa,\uparrow} \hat{c}_{\kappa, \uparrow} \right) \hat{U}^\dag_{\rm LLP:2} \hat{U}^\dag_{\rm LLP}$. Note, as the center of mass coordinate is not guaranteed to belong to the crystal lattice, it is not possible to perform the LLP transformation directly into the center of mass frame. This is why the total transformation has a two-step character, where we first move to the frame of the spin-up electron and then move halfway to the spin-down electron. The final frame corresponds to the center of mass frame and $\hat{c}_{n,\downarrow}$ represents the fermionic annihilation operator corresponding to the relative coordinate of the two electrons.

It is useful to consider how the remaining symmetries of the Hamiltonian manifest themselves in the center of mass frame. Exchanging the two electrons in the initial frame is equivalent to flipping the sign of the relative coordinate, which in operator language amounts to $\hat{c}_{n,\downarrow} \rightarrow e^{i\theta} \hat{c}_{-n,\downarrow}$. Since all the terms in the above Hamiltonian depend only on $n\delta$ product, the symmetry is trivially satisfied. Furthermore, the effective parity symmetry in the $t=0$ limit is simplified. The position of the relative coordinate plus the total boson number is conserved modulo two. The Hilbert space of the final Hamiltonian \eqref{eqn:Hamiltonian_final} for $t=0$ is spanned by two disconnected parts:  even parity $P_+: \{ c^\dag_{2n,\downarrow} \otimes (b^{\dag})^{2m} \ket{0},c^\dag_{2n+1,\downarrow} \otimes (b^{\dag})^{2m+1} \ket{0} \}$ or odd parity $P_-: \{ c^\dag_{2n + 1,\downarrow} \otimes (b^{\dag})^{2m} \ket{0},c^\dag_{2n,\downarrow} \otimes (b^{\dag})^{2m+1} \ket{0} \}$, where $n,m$ are integers. 

The LLP transformation reduces the number of free variables but introduces additional non-local interactions, such as $\cos(\hat{Q}_b - \kappa)$ and $\sin(\hat{Q}_b - \kappa)$. One of the key benefits of this transformation is that it introduces entanglement between multiple degrees of freedom. Thus even simple factorizable states in the LLP frame correspond to highly entangled states in the lab frame. Our next step is to introduce a variational ansatz that has a relatively small number of free parameters but efficiently captures correlations in the system.

\subsection{Non-Gaussian variational ansatz}

We now introduce a variational ansatz in the form that explicitly manifests symmetries of the model and discuss the balance between keeping the number of parameters small while making the ansatz sufficiently expressive. The simplest choice of the variational wave function in the LLP frame is an outer product of a single electron wave function with a coherent state of phonons. Since the phonon number is not conserved, the coherent state is a natural choice, enabling efficient computation of expectation values and keeping the number of variational parameters low. However, this simple ansatz cannot capture the relevant symmetries of the system discussed before. Instead, we consider a generalized version of this ansatz, constructed from the sum of coherent states and single-electron wave functions 
    \begin{equation}
        \ket{\psi} = \sum_c^{N_c} \ket{\alpha_c} \otimes \ket{\Delta_c},
    \end{equation}
    where $\ket{\Delta_c}=e^{i\sum_q \hat{b}_q^\dag \Delta_{c}(q) + \hat{b}_q \Delta^*_{c}(q)} \ket{0}$, $\ket{\alpha_c} = \sum_n \alpha_{n,c} \hat{c}_{n,\downarrow}^\dag \ket{0}$,  $N_c$ is the number of coherent states in the sum and $\ket{0}$ is the vacuum state. For the ansatz to satisfy the \textit{parity} symmetry, relevant in the strong coupling regime, we require at least four coherent states related by the parity symmetry.  To understand this, we first note that the phonon coherent state combination $(\ket{\Delta_c} \pm \ket{-\Delta_c})$ contains an even/odd number of phonons. Therefore, in $t=0$ limit the ansatz would reduce to the minimal form $\ket{\alpha_{odd}} \otimes (\ket{\Delta_c} \mp \ket{-\Delta_c}) + \ket{\alpha_{even}} \otimes (\ket{\Delta_c'} \pm \ket{-\Delta_c'}) $ for even/odd parity $P = \pm 1$. Therefore, the parity symmetry requires the \textit{ansatz} to have a minimum of four coherent states. In addition to the parity symmetry, the system satisfies the time-reversal symmetry $K$ in all the parameter regimes, which doubles the number of necessary coherent states, making the minimum number of necessary coherent states eight. Given all of the constraints, we use an ansatz which by construction satisfies the conjugation symmetry and contains eight independent coherent states, giving a total of 16 effective states 
    \begin{equation}
        \ket{\Psi_{\rm LLP }} = (1 + K) \sum_{c = 1}^{8} \ket{\alpha_c} \otimes \ket{\Delta_c},
        \label{eq:ansatz}
    \end{equation}
    where $K$ is the conjugation symmetry introduced in the previous section. Note, in the strong coupling limit, the number of independent coherent states reduces to two, as each set of four is related by the parity symmetry. Since we use coherent states of phonons, there is no truncation of the phonon basis, allowing the efficient simulation of all parameter regimes. To get a more intuitive picture of the above \textit{ansatz} \eqref{eq:ansatz}, we compute the full expression $\ket{\Psi_\kappa} = \hat{U}_{\rm LLP}  \hat{U}_{\rm LLP:2} \hat{c}_{\kappa, \uparrow}^\dag \ket{\Psi_{\rm LLP}}$ and re-arrange to obtain 
   \begin{equation}
   \begin{split}
        \ket{\Psi_\kappa} = &\sum_j e^{i(\kappa - \hat{P}_{\rm TOT} )j} \times\\
        &\left[ \hat{c}^\dag_{0,\uparrow} \sum_{c = 1}^{8} \sum_n \alpha_{n,c} \hat{c}^\dag_{n,\downarrow} \ket{0} \otimes e^{-i\hat{Q}_bn/2}\ket{\Delta_c}  \right],  
   \end{split}
   \end{equation}
where $\hat{P}_{\rm TOT} = \sum_{k,\sigma} k \hat{c}_{k,\sigma}^\dag \hat{c}_{k,\sigma} + \hat{Q}_b$ is the total system momentum. The total wave function is made up of placing one electron at site zero and the second one in an orbital around it, along with the appropriate phonon cloud, based on the center of mass position. This local object is then translated in real space with the appropriate phase factor to obtain the overall translationally invariant wave function with a definite momentum $\kappa$. 

\subsection{Variational dynamics and the excitation spectrum} \label{sec:tangent-space}

To obtain the ground and excited states of the SSH Hamiltonian \eqref{eqn:Hamiltonian_final}, we follow the formalism developed for a general variational ansatz ~\cite{shi_variational_2018,hackl_geometry_2020}. The main idea is to restrict the usual real/imaginary time evolution to the variational manifold. When the class of variational wavefunctions is chosen to be sufficiently expressive, this projected dynamics should be close to the exact evolution in the full Hilbert space. The general form of imaginary-time evolution is given by the equation

\begin{equation}
    \frac{d}{d\tau} \ket{\psi(\tau)} = - (\hat{H} - E(\tau)) \ket{\psi(\tau)},
\end{equation}
where $\tau = it$ is the imaginary time and $E(\tau) = \bra{\psi(\tau)}\hat{H}\ket{\psi(\tau)}$. For an ansatz parametrized by a set of parameters $\{x^\mu \}$, the projection of the evolution equation onto the variational subspace gives  

\begin{equation}
    \frac{d x^\mu}{d\tau} = - \sum_\nu G^{\mu \nu} \partial_\nu E,
    \label{eqn:minimization}
\end{equation}
where $\partial_\mu \equiv \frac{\partial}{\partial x^\mu}$ and we introduced the Gram matrix $g_{\mu,\nu} = \Re{\bra{\partial_\mu \psi} Q_\psi \ket{\partial_\nu \psi}}$, with $Q_\psi = \mathds{1} - \ket{\psi}\bra{\psi}$. The Gram matrix defines a natural metric on the tangent space, where the current state $\ket{\psi}$ is projected out and $G^{\mu,\nu}$ is its inverse, such that $\sum_\nu G^{\mu,\nu} g_{\nu,\rho} = \delta^\mu_{\rho}$. Starting from a random initial state, we evolve the parameters according to \eqref{eqn:minimization} or using gradient descent methods, giving a set $\{\Bar{x}^\mu\}$, which minimizes the energy of the system.  For the case of variational ansatz \eqref{eq:ansatz}, the parameter space consists of electronic orbitals and phononic coherent states $\{x^\mu \} = \{ \Re{\alpha_{n,c}}, \Im{\alpha_{n,c}}, \Re{\Delta_{c}(q)}, \Im{\Delta_{c}(q)}\}$. For details on how to evaluate the energy $\bra{\Psi_\kappa} \hat{H} \ket{\Psi_\kappa}$ and the tangent vectors $ \ket{ \partial_{\alpha_{nc}/\Delta_c(q)} \Psi_{\kappa}}$, see Appendix \ref{sec:expec}.

To obtain the excited states and the spectrum, there are two complementary approaches ~\cite{hackl_geometry_2020,shi_variational_2018}. Either diagonalize the Hamiltonian in the tangent space or linearize the equations of motion around the ground state. In general, these can give different results and here we choose to follow the former approach. It can be shown ~\cite{hackl_geometry_2020} that the $n$-th computed eigenvalue $\omega_n$ is an upper bound to the true n-th eigenstate energy $E_n$, $\omega_n \geq E_n$. The projected tangent space Hamiltonian is
\begin{equation}
    H^\mu_\nu = \sum_\rho G^{\mu\rho} \Re{\bra{V_\rho} \hat{H} \ket{V_{\nu}}},
\end{equation}
where $\ket{V_\mu} = Q_\psi \ket{\partial_\mu \psi}$ is the projected tangent space vector. By diagonalizing the Hamiltonian matrix $H^{\mu}_{\nu}$, we obtain a set of coefficients $X^\mu_{n}$, where the $n$-th eigenvector is $\ket{X_n} = \sum_\mu X^\mu_{n} \ket{V_\mu}$. Note, due to the K\"ahler property of the tangent space, the above equation gives rise to double degeneracy of every energy level ~\cite{hackl_geometry_2020}. For any vector $\ket{V_\mu}$ in the tangent space, $i\ket{V_\mu}$ is a distinct vector also in the tangent space, having the same energy. In the following, we neglect this degeneracy, since it does not affect any observables. As the tangent space is formed by only linear excitations around the ground state, we focus our studies on the low energy spectrum. In the case of ansatz \eqref{eq:ansatz}, the tangent space consists of combinations of particle-hole and single phonon excitations on top of the coherent state sum.

Given the excitation spectrum, we can also compute the dynamical response function $\Im{\chi_{\rm O}(q,\omega)}$, with $\hat{O}$ being any operator. Generally, the response can be obtained using Lehmann representation~\cite{dupuis2023field}, which for zero temperature and $\omega > 0$ can be written as

\begin{equation}
    \Im{\chi_{\rm O}(q,\omega)} = -\pi \sum_n |\bra{n} \hat{O}(q) \ket{GS}|^2 \delta(\omega - \omega_{n}),
    \label{eqn:response}
\end{equation}
where $\ket{GS}$ is the ground state, $\ket{n}$ is the $n$-th excited state of the system and $\omega_{n} = E_n - E_{GS}$. The usual complexity of the above equation is knowing the exact eigenstates of the system, which lead to the development of many approximate methods, including the linearized equations of motion around the variational ground state ~\cite{hackl_geometry_2020}. In this work, we instead use the approximate excited states $\ket{X_n}$ and evaluate the response \eqref{eqn:response} directly.

\section{Bipolaron ground state}\label{sec:GS}

    In this section, we study the ground state properties of the SSH bipolaron using the summed coherent state wave function \eqref{eq:ansatz} for a 1D chain with $N = 100$ sites. Depending on the relative phonon and electron energies, there are two main parameter regimes: $\Omega = 0.5 t $ (\textit{adiabatic}) and $\Omega = 3.0 t $ (\textit{anti-adiabatic}) ~\cite{sous_light_2018}. In these two regimes, we sweep the dimensionless electron-phonon interaction strength $\lambda = \frac{2g^2}{\Omega t} $ from weak ($\lambda \ll 1$) to strong coupling limit ($\lambda \sim 1 $).  Using gradient descent and imaginary time evolution, we minimize the energy $E(\kappa) = \bra{\Psi_\kappa} \hat{H} \ket{\Psi_\kappa}$  w.r.t. variational parameters and obtain the ground state $\ket{\Psi_\kappa\{ \alpha_{n,c}, \Delta_{c,q} \}}$.
    
    \subsection{Dispersion and electron momentum distribution}

\begin{figure}
        \centering
        \includegraphics[width=1.0\linewidth]{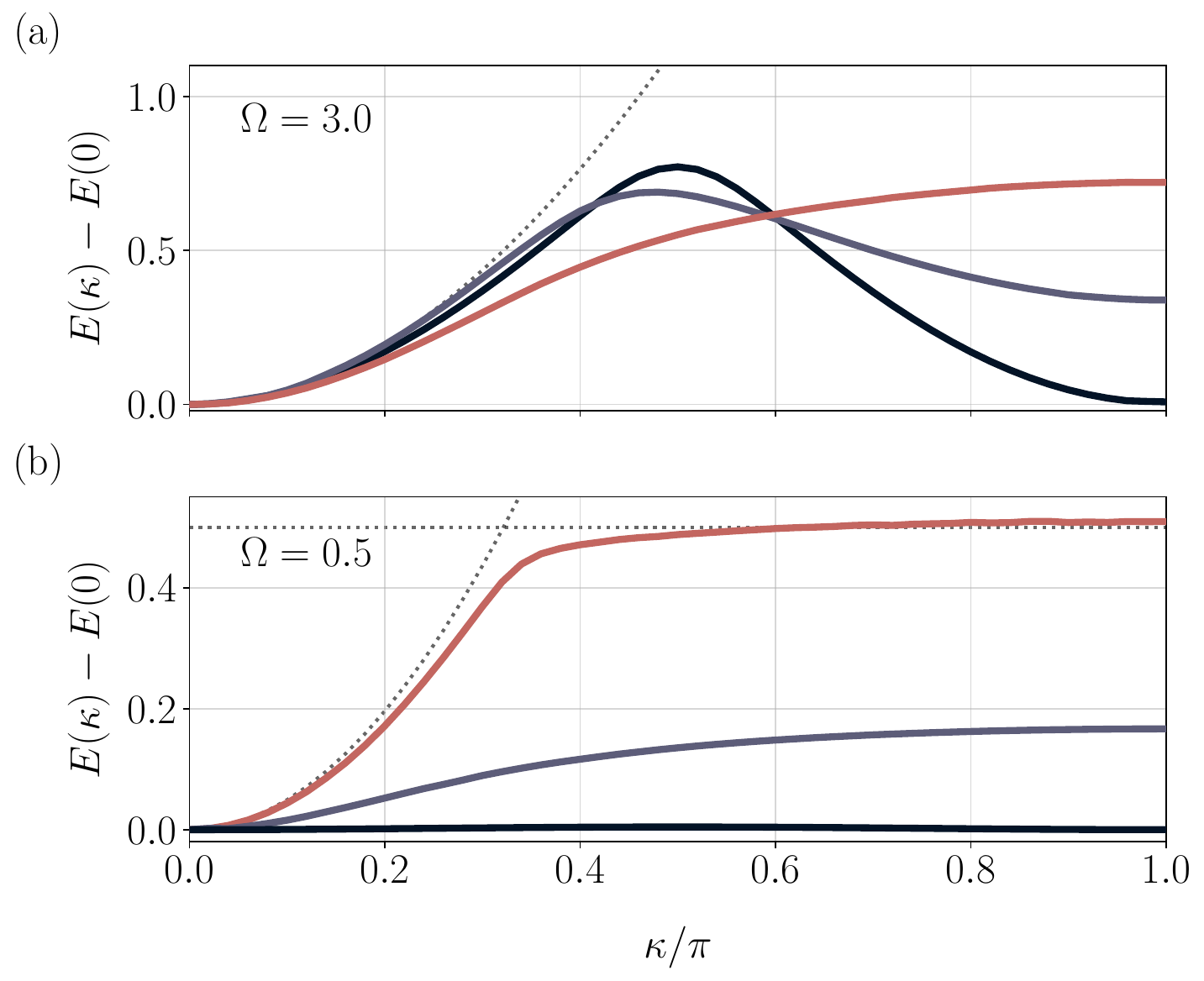}
        \caption{Dispersion $E(\kappa) - E(\kappa = 0)$ in units of hopping $t$, as a function of total momentum $\kappa$ for different values of electron-phonon coupling. a) Energy dispersion in the anti-adiabatic regime $\Omega = 3.0t$,  for different values of $\lambda$:  $0.5$(\textcolor{red1}{\sampleline{}}), $0.7$(\textcolor{purple2}{\sampleline{}}) and $2.0$ (\textcolor{black1}{\sampleline{}}). The dotted line indicates the free fermion dispersion. b)  Energy dispersion in the adiabatic regime $\Omega = 0.5t$, for different values of $\lambda$:  $0.15$(\textcolor{red1}{\sampleline{}}), $0.5$(\textcolor{purple2}{\sampleline{}}) and $0.8$ (\textcolor{black1}{\sampleline{}}). The dotted line indicates the free fermion dispersion and the horizontal line indicates the single phonon excitation energy $E(0) + \Omega$. }
        \label{fig:energy-dispersion}
    \end{figure}

    Figure \ref{fig:energy-dispersion} shows the dispersion for different values of coupling in the adiabatic and the anti-adiabatic regimes. Even for small couplings, in the adiabatic regime, the dispersion is significantly modified at large momenta (see Figure \ref{fig:energy-dispersion}b, $\lambda = 0.15$). Strong renormalization of the dispersion can be understood from the fact that at large momenta it is energetically favorable to transfer the system momentum to a single phonon, which only increases energy by $\Omega$. This mechanism is particularly effective in the adiabatic regime, where the phonon energy is smaller than the hopping, resulting in characteristic flat dispersion at higher values of the total momentum. For higher coupling, the dispersion develops a non-monotonic behavior, going from singly periodic at $\lambda = 0$, to almost doubly periodic at $\lambda \gg 1$. Even though phonons are localized relative to the electrons, the effective lattice for the bipolaron appears almost perfectly dimerized  (cf. half-filled SSH model ~\cite{su_solitons_1979}). For all values of coupling, the ground state remains at zero momentum, in contrast with the single polaron result, where the ground state gets shifted from $0$ to $\pm \pi/2$ for large coupling ~\cite{marchand_sharp_2010,shi_variational_2018} (see Appendix \ref{sec:figures}). We point out the difference in dispersions at strong coupling for the adiabatic and anti-adiabatic regimes. In the former case, the dispersion becomes very flat, before looking as if the system periodicity has been doubled, whereas in the latter case, the dispersion only displays the doubling, without becoming flat.

\begin{figure}
        \centering
        \includegraphics[width=1.0\linewidth]{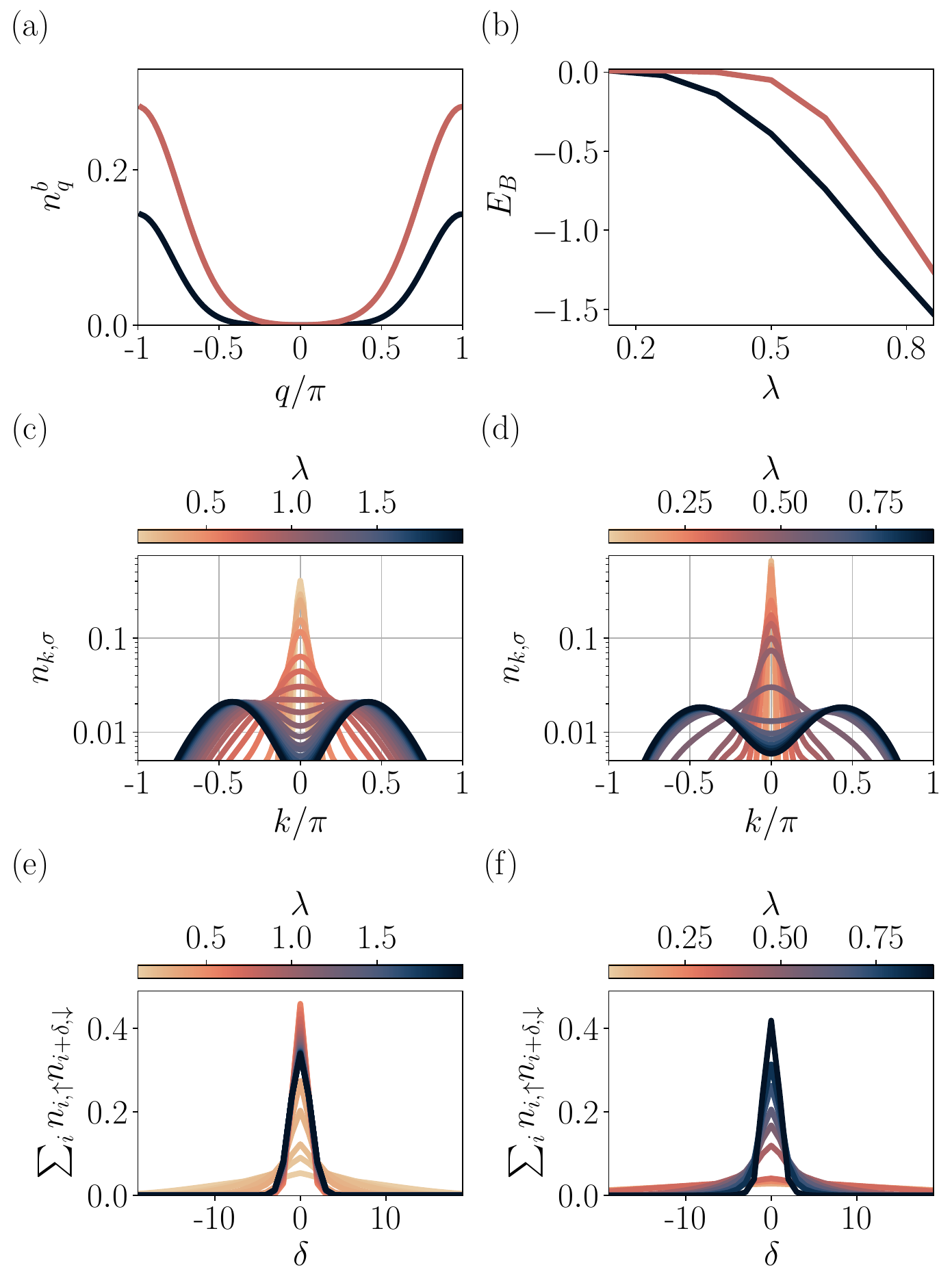}
        \caption{a) Momentum distribution of phonons in the high coupling regime for $\lambda = 2$, $\Omega = 3t$ (\textcolor{black1}{\sampleline{}}) and $\lambda = 0.8$, $\Omega = 0.5t$ (\textcolor{red2}{\sampleline{}}). b)  Binding energy of a bipolaron $E_B = E_{GS}^{BP} - 2E_{GS}^{P}$, in units of $t$, as a function of total coupling $\lambda$ for adiabatic $\Omega = 0.5t$ (\textcolor{red2}{\sampleline{}}) and anti-adiabatic $\Omega = 3t$ (\textcolor{black1}{\sampleline{}}) regime. c)  Bipolaron correlation function $\sum_i \expval{n_{i,\uparrow}n_{i+\delta,\downarrow}}$ for different values of coupling $\lambda$ at $\kappa =0$ for $\Omega =3t$. d) Bipolaron correlation function $\sum_i \expval{n_{i,\uparrow}n_{i+\delta,\downarrow}}$ for different values of coupling $\lambda$ at total momentum $\kappa =0$ for $\Omega = 0.5t$.}
        \label{fig:binding-correlations}
    \end{figure}

    The effective lattice dimerization can be understood from the form of the electron-phonon coupling Hamiltonian \eqref{eqn:SSH-coupling}. The phonon dispersion is flat and the electrons couple to the difference of phonon displacement between two sites $(x_{n+1} - {x_n})$. The difference is maximized for the neighboring sites with an opposite sign, corresponding to $\pi$ momentum phonons. However, it is only beneficial to displace the lattice in the vicinity of the electrons, resulting in a locally dimerized lattice, which corresponds to a momentum distribution of phonons $\expval{\hat{n}_q^b} = \bra{\Psi_{\rm LLP}} \sum_q \hat{b}^\dag_q \hat{b}_q \ket{\Psi_{\rm LLP}} $ centered around the $\pm \pi$ momenta, with the width governed by the bipolaron size (see Figure \ref{fig:binding-correlations}a). 
    To understand the dimerization in more detail, we consider the single-electron momentum distribution  $\expval{\hat{n}_{k,\sigma}} = \bra{\Psi_\kappa} \hat{c}_{k,\sigma}^\dag \hat{c}_{k,\sigma} \ket{\Psi_\kappa} = \bra{\Psi_{\rm LLP}} \sum_{r,n} \hat{c}_{n-r,\downarrow}^\dag \hat{c}_{n,\downarrow} e^{i(k - \kappa/2 - \hat{Q}_b/2)r} \ket{\Psi_{\rm LLP}} $. As shown in Figure \ref{fig:binding-correlations}c-d, at small couplings the individual electrons occupy states around the zero momentum, due to kinetic energy costs. With increasing coupling, the electrons shift their momentum distribution to be around $ k = \pm \pi/2$. This is equivalent to the single polaron SSH problem, where at strong coupling the minimum of polaron dispersion corresponds to the $\pi/2$ momentum state~\cite{marchand_sharp_2010} (see Appendix \ref{sec:figures}). The main difference is that in the bipolaron case, the electrons contain an additional degree of freedom corresponding to the relative momentum.
    We can now understand the periodicity doubling of bipolaron dispersion, from the perspective of the electrons. At strong coupling, the bipolaron is formed from two polarons, with individual momentum distributions localized around $\pm \pi/2$. Therefore, the low energy combination of the two polarons can form either a total momentum $\kappa = 0 $ or $\kappa = \pi$ state, resulting in a dispersion that resembles lattice periodicity doubling. With finite hopping $t$, the momentum distributions are skewed towards zero momentum, so the individual polaron momenta can cancel to give $\kappa = 0$, but cannot add exactly to $\kappa = \pi$, breaking the double periodicity and keeping the ground state at zero momentum, as favored by bare hopping. 
    The unconventional change in the momentum distribution of electrons is a characteristic property of the momentum-dependent interaction vertex $g(k,q)$. Given the form $g(k,q) \propto \sin(k) - \sin(k+q)$, the interaction strength is maximized for electrons with momenta close to $\pi/2$ and phonons with momenta close to $\pi$. Therefore, at large couplings, the electrons distribute around a finite momentum value. In Holstein-like models, the vertex strength is independent of electron momentum, so the electron distribution is dictated by the kinetic energy and the electrons predominantly occupy states with momenta close to zero. The shift of electron momentum distributions is significant as it demonstrates new phenomenology. For small coupling, the distribution is centered around the zero momentum, such that the bare electron hopping contributes significantly to the total energy, while the coupling to phonons does not contribute significantly. At high coupling, the momentum distribution is centered around $\pi/2$, demonstrating that phonon-assisted hopping is now dominant. Interestingly, we find that when one type of hopping mechanism (phonon independent or phonon facilitated) becomes dominant, the other one gets suppressed. Competition between dominant mechanisms of electron hopping underlies the phase transition in the single polaron problem~\cite{marchand_sharp_2010,PhysRevB.89.144508}. In that case the state with minimum energy shifts from zero to finite momentum with increasing interaction strength. This behavior indicates that in systems with momentum-dependent electron-phonon coupling, there is an additional mechanism of suppressing polaron propagation, which differs from the standard argument of the phonon clouds overlap ~\cite{alexandrov_polaron_2000, devreese_frohlich_2009, devreese_frohlich_2015}, explained in detail later in Sec \ref{sec:effective-mass}. In the next section, we will see this mechanism has a strong effect on the effective mass and gives rise to an emergent parity symmetry in the strong coupling regime, where the bare hopping is suppressed.

    \subsection{Binding energy and bipolaron radius}
    
    As suggested by the electron momentum distribution, the bipolaron can be viewed as a composite of two polarons, bound by the phonon cloud. The individual electrons distort the lattice to balance the energy cost of creating phonons and energy gain from electron-phonon interaction, forming a polaron. Sharing the phonon cloud allows for the same energy benefit due to coupling, while requiring the creation of fewer phonons in total, binding the two polarons into a bipolaron. Figure \ref{fig:binding-correlations}e-f shows the electronic density correlation  $ \expval{\sum_i \hat{n}_{i,\uparrow} \hat{n}_{i+\delta,\downarrow}} = \bra{\Psi_\kappa} \sum_i \hat{c}^\dag_{i,\uparrow} \hat{c}_{i,\uparrow} \hat{c}^\dag_{i+\delta,\downarrow} \hat{c}_{i+\delta,\downarrow} \ket{\Psi_\kappa}  = \bra{\Psi_{\rm LLP}} \hat{c}^\dag_{\delta,\downarrow} \hat{c}_{\delta,\downarrow} \ket{\Psi_{\rm LLP}}$ as a function of the coupling in both regimes. Already at small coupling, the wave function in the LLP frame becomes localized, which in the original frame, corresponds to the two polarons forming a bound state. This is consistent with the usual expectation that any attractive potential in 1D leads to a bound state, even though in the case of the SSH model, the effective electron-electron interaction does not reduce to the usual density-density form ~\cite{sous_light_2018}. Increasing the coupling changes the localization range (i.e. the radius of bipolaron) non-monotonically, reaching the maximum at an intermediate value. This can be directly seen in Figure \ref{fig:binding-correlations}e-f, where the correlation function shows the highest peak for intermediate coupling. 
    The non-monotonic behavior can be associated with shifting between the two different hopping mechanisms, bare or phonon-assisted.  
    Figure \ref{fig:binding-correlations}b shows the binding energy $E_B$ as a function of coupling for both regimes. We define the binding energy as the difference between the bipolaron ground state and the ground state energy of two polarons $E_B = E_{GS}^{BP} - 2 E_{GS}^P$, where the polaron ground state was computed using similar methods as for the bipolaron (see Appendix \ref{sec:figures}). For coupling values close to zero, the binding energy falls below the numerical precision of the variational methods, however, in this regime, the perturbative analysis ~\cite{sous_light_2018} has shown that the bipolaron is bound. As the coupling is increased, the binding energy increases and the bipolaron remains bound for all values of the coupling, in agreement with the correlation function results discussed previously.
    
    \subsection{Effective mass} \label{sec:effective-mass}

    \begin{figure}
        \centering
        \includegraphics[width=1.0\linewidth]{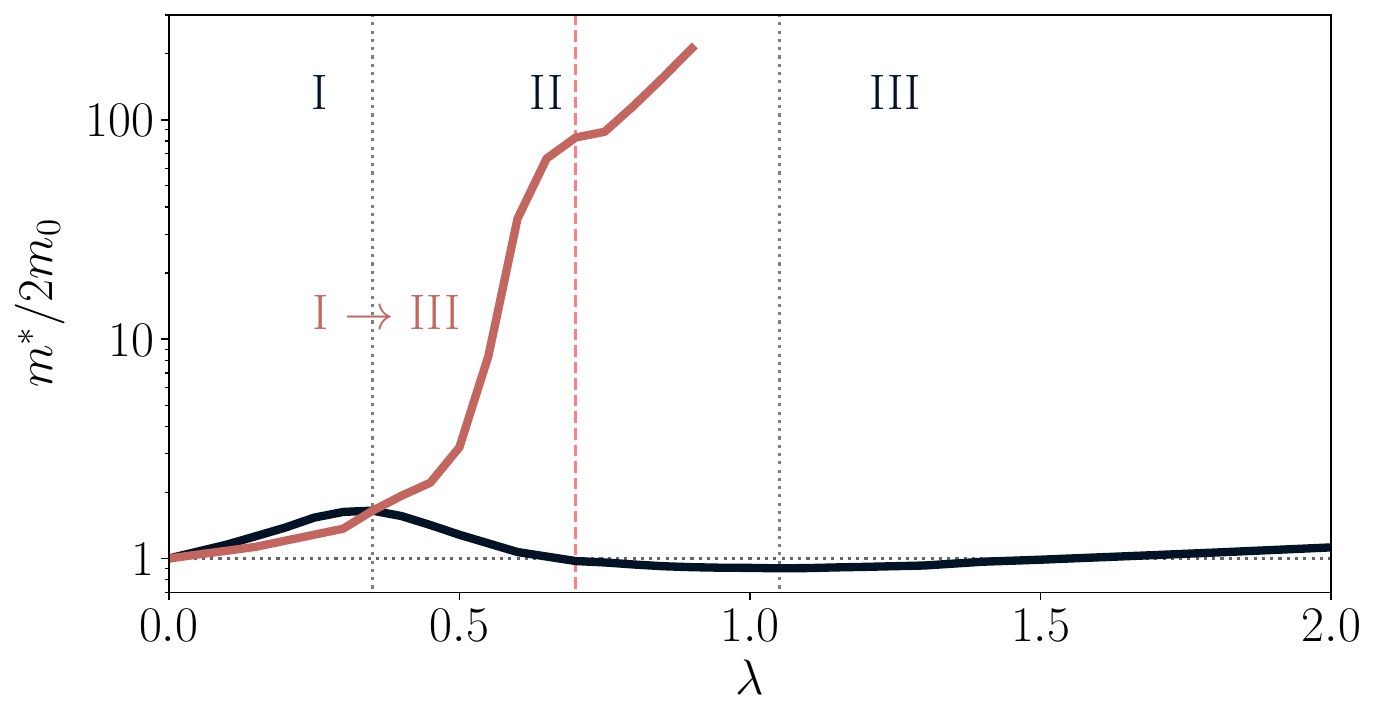}
        \caption{Effective mass $m^*$ of the bipolaron as a function of coupling $\lambda$ for adiabatic $\Omega = 0.5t$ (\textcolor{red2}{\sampleline{}}) and anti-adiabatic $\Omega = 3t$ (\textcolor{black1}{\sampleline{}}) regimes. The anti-adiabatic regime shows three distinct parameter regions $\rm I - II - III$, as indicated by the gray dotted lines, corresponding to three different behaviors of the effective mass. The adiabatic regime transitions directly from regime $\rm I $ to $\rm III$,
        with the red dashed line indicating the transition between bare and phonon-mediated hopping.} 
        \label{fig:effectivemass}
    \end{figure}
    
    Having established that the bipolaron is bound for every value of coupling, we compute its effective mass  $m^* = 1/ \frac{\partial^2 E_{\rm GS}}{(\partial \kappa)^2}$. As discussed previously for the anti-adiabatic regime ~\cite{sous_light_2018} and confirmed by our results (see Figure \ref{fig:effectivemass}), the bipolaron remains light for all the values of the electron-phonon coupling considered. This is in contrast with the Holstein model, where the effective mass grows exponentially with the coupling strength ~\cite{alexandrov_advances_2010}. The exponential increase can be understood by considering the polaron hopping process, which involves the movement of the phonon cloud. During that process, the hopping gets suppressed by phonon cloud overlap, which is exponentially small in the total phonon number and therefore in the coupling strength as well. In the case of the SSH model, the increase in coupling also enhances hopping strength between the sites (through $g$ and $\expval{x}$), so the cloud overlap argument does not apply straightforwardly. Based on the behavior of the effective mass, there are three different regimes in the anti-adiabatic limit (see Figure \ref{fig:effectivemass}). 
    In the first regime (perturbative in $g$), higher coupling implies that electron hopping moves more phonons. Since the electrons are still localized around $k = 0$, they do not benefit from the phonon-assisted hopping, and the bipolaron just becomes heavier. In the second regime ($\lambda \sim 1 $), there is a transition and the hopping becomes dominated by phonon-assisted processes. The electrons redistribute around  $k = \pi/2$ momenta, as shown in the electron momentum distribution (see Figure \ref{fig:binding-correlations}c). As the coupling increases, the effective hopping of the bipolaron increases, lowering the effective mass. Since the electron momentum distribution evolves continuously, we expect no discontinuities in the effective mass as a function of coupling between the two hopping regimes.
    In the third regime ($g \gtrsim \Omega$), the number of phonons is high, due to high coupling. A competition then arises between the Holstein-like, phonon orbital suppression and the hopping enhancement of the $g\expval{x}$ term. As a result, the effective mass changes non-monotonically. Initially, the hopping benefits from increasing the coupling (second regime), however at large couplings, the exponential suppression dominates, causing the increase of effective mass. For the anti-adiabatic regime with $\Omega = 3t$ and couplings within the range considered $\lambda < 2$, the full exponential suppression does not dominate and the bipolaron remains light. 
    The adiabatic regime is in complete contrast with this behavior and within the range of couplings considered, the effective mass increases up to $200 \times (2m_0)$, where $m_0$ is the bare electron mass. The main difference between the adiabatic and the anti-adiabatic regimes is the relative size of $\Omega/t$. Therefore,  by increasing the coupling $g$, the system first enters the regime $g \gtrsim \Omega$ and then $\lambda \sim 1$, so the second regime, where the effective mass decreases, is absent.  The adiabatic polaron cannot benefit from the phonon-mediated hopping mechanism (regime II) to reduce the mass. Instead, the polaron immediately becomes exponentially heavy due to the phonon dressing of the bare hopping. In that sense, the adiabatic regime bears similarities with the Holstein model, as the bare hopping gets re-normalized due to phonon cloud overlap. Once the coupling is high enough and $\lambda \sim 1$, the phonon-mediated hopping becomes relevant, however, the number of phonons is so large, that the exponential hopping suppression dominates and the effective mass keeps increasing. The transition between the two hopping mechanisms results in a change of effective mass growth rate, as expected, and demonstrates similar saturation behavior as the anti-adiabatic regime I to regime II crossover. In conclusion, for momentum-dependent electron-phonon vertex $g(k,q)$, both the $\Omega /t$ and $\lambda$ energy scales are important. Depending on the relative size of these scales, qualitatively different regimes can arise, leading to orders of magnitude differences in effective mass behavior.
    
    \begin{figure*}
        \centering
        \includegraphics[width=1.0\linewidth]{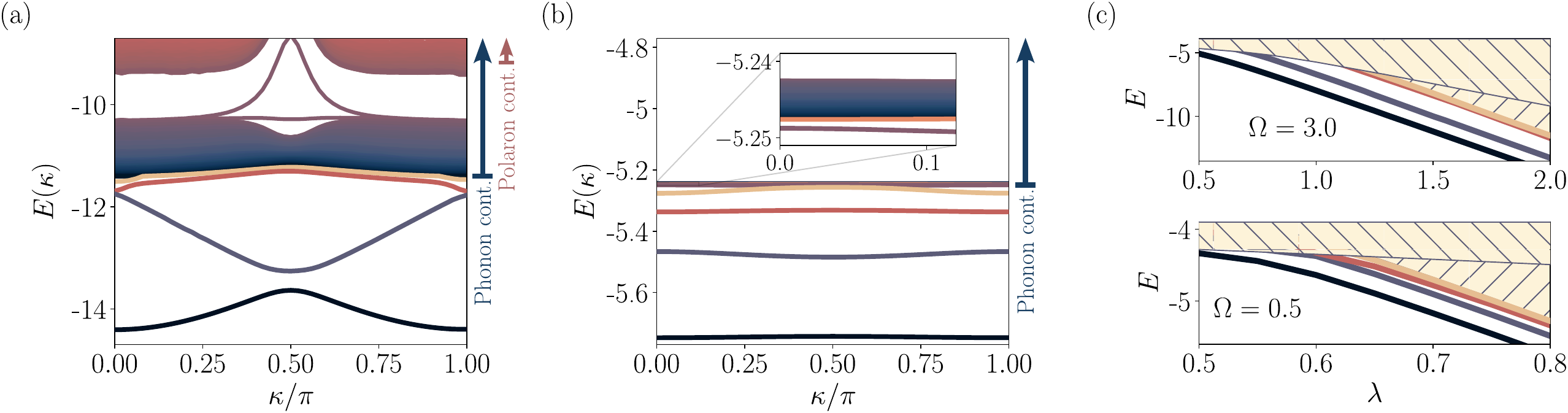}
        \caption{The excitation spectrum of the bipolaron. a) Band dispersion $E(\kappa)$, in units of $t$, in the anti-adiabatic regime $\Omega = 3t$ with $\lambda = 2$. The arrows on the side indicate the energy range of the single phonon (blue) and the unbound polaron continuum (red) b) Band dispersion $E(\kappa)$, in units of $t$, in the adiabatic regime $\Omega = 0.5t$ with $\lambda = 0.8$. The blue arrow indicates the energy range of the phonon continuum. The inset shows the fourth (\textcolor{purple3}{\sampleline{}}) and fifth (\textcolor{red3}{\sampleline{}}) excited band with very small binding energies. c) Evolution of the four lowest band dispersion as a function of coupling $\lambda$ in the anti-adiabatic (top) and adiabatic (bottom) regime. The width indicates the bandwidth of individual bands.  The blue hatched area represents the single phonon excitation on top of the ground state $E = E_{GS}(\kappa = 0) + \Omega$ and the red hatched area represents two polaron continuum $E = 2 E_{GS}^P$.}
        \label{fig:excitation-spectrum}
    \end{figure*}

\section{Bipolaron excited states}\label{sec:EXC}

In this section, we study the excitation spectrum of the SSH bipolaron by constructing linear excitations around the ground state (see Sec \ref{sec:tangent-space}), with the focus on the strong coupling limit of the adiabatic ($\lambda = 0.8,\Omega = 0.5t$), and the anti-adiabatic ($\lambda = 2, \Omega = 3t$) regimes (see Figure \ref{fig:excitation-spectrum}a-b). We show the spectrum consists of a set of discrete bands, a single-phonon continuum, and a mixed particle-hole and phonon continuum. The discrete bands correspond to bound bipolaron-phonon states, arising from the deformation of the phonon cloud around the bipolaron. In the anti-adiabatic regime, the discrete bands have a well-defined parity, as the bare hopping is suppressed. These bands always appear in pairs, one band corresponding to $P_+ $ and the other band corresponding to $P_-$ sector. We conclude by discussing how the strong coupling properties change as a function of the coupling strength (see Figure \ref{fig:excitation-spectrum}c) and comparing our results to the Holstein model. 

\subsection{General properties of the spectrum}

Before describing the computed spectrum, we first discuss the general properties of the system in the thermodynamic limit. Given the form of the Hamiltonian, the excitation spectrum contains a single-phonon continuum, consisting of a bipolaron and an unbound phonon, and a polaron continuum, consisting of two unbound polarons. Due to the coupling of electron and phononic degrees of freedom, the spectrum also contains a mixed particle-hole and phonon continuum. The lower-energy edge of the single-phonon continuum, defined over the entire momentum range is $E_{GS}(\kappa = 0) + \Omega$, as any finite momentum of the system can be transferred to the phonon. Similarly, the polaron continuum corresponds to energies above the ground state of two polarons $2E_{GS}^P$, with the polaron dispersion shown in the Appendix \ref{sec:figures}. The arrows in Figure \ref{fig:excitation-spectrum}a-b indicate the energy range corresponding to the two continua. The high energy required to dissociate a bipolaron into two unbound polarons directly reflects the stability of the bipolaron in the strong coupling regime. Any state in the spectrum with energy below the two continua corresponds to a bound bipolaron state. This is analogous to a locally interacting two-particle problem. States lying below the continuum do not have the energy required to separate the two particles, making the states bound. Note in some cases, at finite values of the total momentum, states with energies above the lower edge of the two continua can still be bound, as the simple energy threshold $E_{GS}(\kappa = 0) + \Omega$, introduced above, is not exact. Due to the coupling of a finite momentum phonon and the bipolaron, the lower edge of the phonon continuum is lifted in energy, compared to the simple thermodynamic arguments given before. Therefore, the low energy edge of the single-phonon continuum $E_{GS}(\kappa = 0) + \Omega$, can only be used as a sufficient condition to indicate states below the threshold are bound, while the states above the threshold need to be further characterized.

\subsection{First excited state} \label{sec:first-exc}

\begin{figure}
        \centering
        \includegraphics[width=1.0\linewidth]{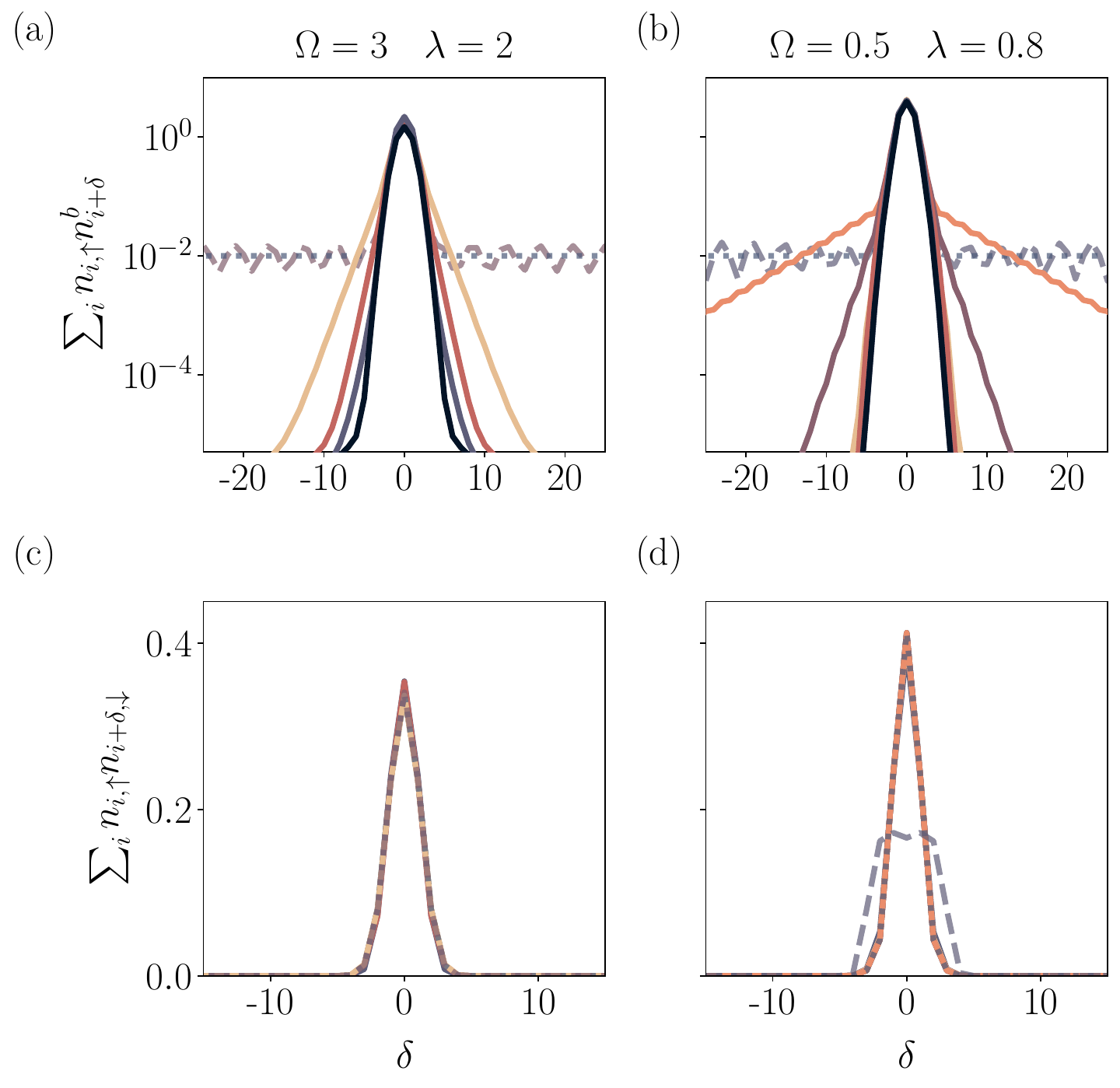}
        \caption{a) Electron-phonon correlation function $\expval{\sum_i n_{i,\uparrow} n^b_{i+\delta}}$ in the anti-adiabatic regime $\Omega =3t$ with $\lambda =2$ at zero total momentum $\kappa = 0$, for the ground state (\textcolor{black1}{\sampleline{}}),
        the three bound excitations ((\textcolor{purple2}{\sampleline{}}),(\textcolor{red2}{\sampleline{}}),(\textcolor{yellow1}{\sampleline{}})) and the representative states of the single-phonon (\textcolor{purple4}{\sampleline{dotted}}) and the mixed particle-hole and phonon continuum (\textcolor{purple5}{\sampleline{dashed}}). 
        b)  Electron-phonon correlation function $\expval{\sum_i n_{i,\uparrow} n^b_{i+\delta}}$ in the adiabatic regime $\Omega =0.5t$ with $\lambda =0.8$ at zero total momentum $\kappa = 0$, for the ground state (\textcolor{black1}{\sampleline{}}),
        the five bound excitations ((\textcolor{purple2}{\sampleline{}}),(\textcolor{red2}{\sampleline{}}),(\textcolor{yellow1}{\sampleline{}}),(\textcolor{purple3}{\sampleline{}}),(\textcolor{red3}{\sampleline{}})) and the representative states of the single-phonon (\textcolor{purple4}{\sampleline{dotted}}) and the mixed particle-hole and phonon continuum (\textcolor{purple5}{\sampleline{dashed}}). 
         c) Electron-electron correlation function in the anti-adiabatic regime $\Omega =3t$ with $\lambda =2$ at zero total momentum $\kappa = 0$, for the ground state, discrete bands and representative states from the single-phonon and the mixed particle-hole phonon continuum. Colors are the same as in a). d) Electron-electron correlation function in the adiabatic regime $\Omega =0.5t$ with $\lambda =0.8$ at zero total momentum $\kappa = 0$, for the ground state, discrete bands and representative states from the single-phonon and the mixed particle-hole phonon continuum. Colors are the same as in b).}
        \label{fig:exc-correlations} 
    \end{figure}

In this section, we study the properties of the first excited band, in both adiabatic and anti-adiabatic regimes. At strong coupling, the energy of the whole band lies far below the single-phonon continuum (see Figure \ref{fig:excitation-spectrum}a-b), indicating the state is strongly bound. The precise nature of the bound state can be understood by considering the electron-electron correlation function and the electron-phonon correlation function $\expval{\sum_i \hat{n}_{i,\uparrow} \hat{n}^b_{i+\delta}} = \sum_n \bra{\Psi_{\rm LLP}} \hat{c}^\dag_{n,\downarrow} \hat{c}_{n,\downarrow} \hat{b}^\dag_{\delta+n/2} \hat{b}_{\delta + n/2} \ket{\Psi_{\rm LLP}}$, where $\hat{b}_{\delta + n/2}$ is defined through the Fourier transformation of $\hat{b}_q$. Both correlation functions decay at large distances from the bipolaron, further confirming the state is a bound bipolaron state (see Figure \ref{fig:exc-correlations}a-d). Since only the electron-phonon correlation changes significantly compared to the ground state, the first excited state can be understood as a bound state of a bipolaron and a phonon, corresponding to phonon cloud deformations on top of the ground state bipolaron. 

\begin{figure}[h]
    \centering
    \includegraphics[width = 1.0\linewidth]{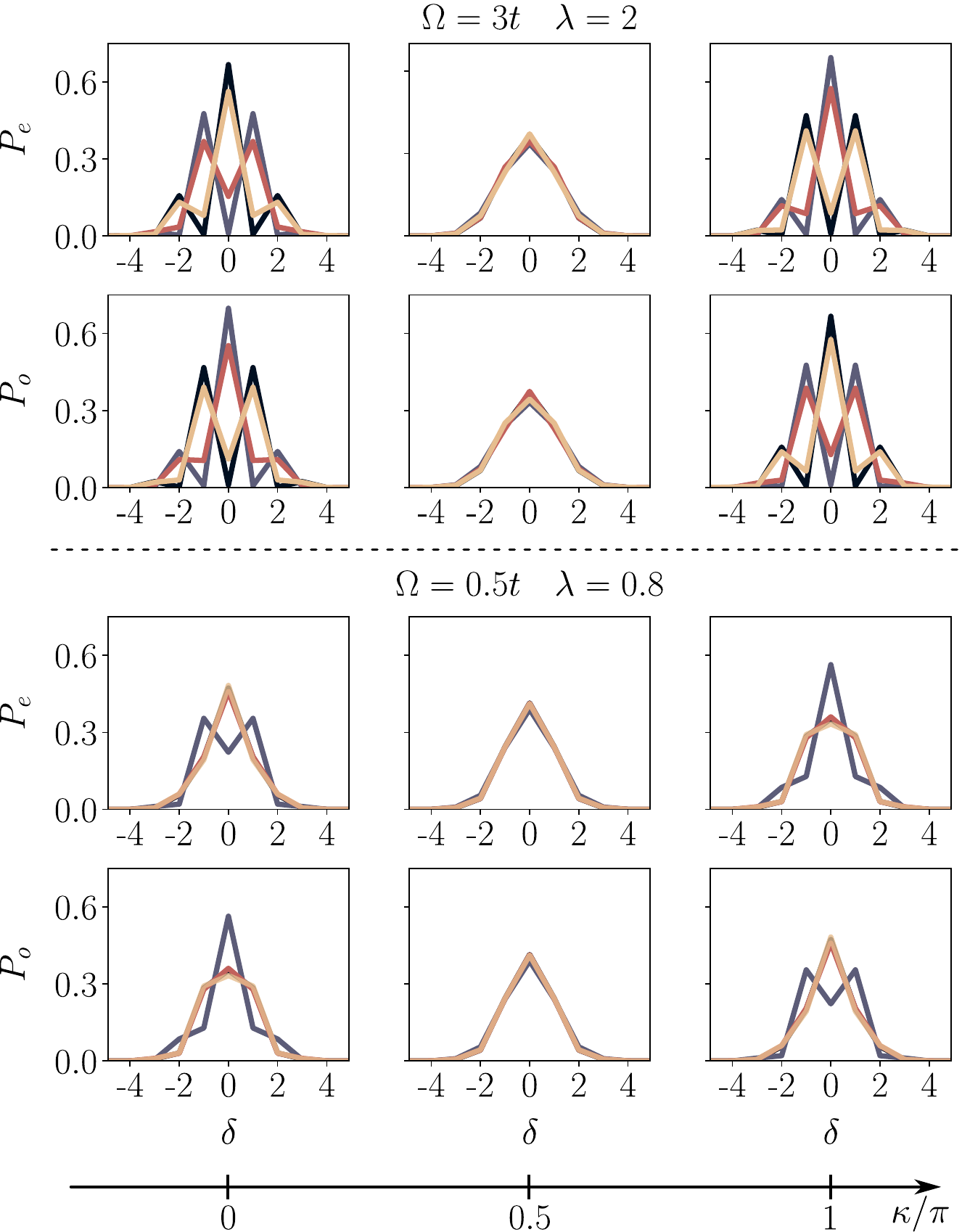}
    \caption{Phonon parity operator $\expval{ \sum_i \hat{n}_{i,\uparrow} \hat{n}_{i+\delta,\downarrow} \hat{P}^b_{\rm o/e}}$, for the ground state (\textcolor{black1}{\sampleline{}}) and first three excited bands ((\textcolor{purple2}{\sampleline{}}),(\textcolor{red2}{\sampleline{}}),(\textcolor{yellow1}{\sampleline{}}), respectively), as a function of electron separation $\delta$. The three columns correspond to the total momentum $\kappa = 0, \pi/2, \pi$ respectively. The upper (lower) panel shows the  anti-adiabatic $\Omega = 3t, \lambda = 2$ (adiabatic $\Omega = 0.5t, \lambda = 0.8$) regime.   
    In the anti-adiabatic regime, at $\kappa = 0$ the ground state shows an even number of phonons for even distances and an odd number of phonons for odd distances, corresponding to even parity $P_+$. The first excited state and the second state show opposite behavior, while the third excited state shows similar structure as the ground state. As the momentum is increased, the parity flips to the opposite value at $\kappa = \pi$ for all the bands. In the adiabatic regime, the parity is not a good quantum number, therefore there is no structure for any momentum value. }
    \label{fig:parity}
\end{figure}

To further understand the origin of the first excited state, we study the parity symmetry of the whole spectrum (see Figure \ref{fig:parity}), by defining a parity operator  
\begin{equation}
    \expval{ \sum_i \hat{n}_{i,\uparrow} \hat{n}_{i+\delta,\downarrow} \hat{P}^b_{\rm o/e}} = \bra{\Psi_{\rm LLP}} \hat{n}_{\delta,\downarrow} \hat{P}^b_{\rm o/e} \ket{\Psi_{\rm LLP}},
    \label{eqn:parity_operator}
\end{equation}
where $\hat{P}^b_{\rm o/e} = 1 \mp (-1)^{\hat{n}^b}$ and $\hat{n}^b = \sum_q \hat{b}^\dag_q \hat{b}_q$. The parity operator determines whether the state has an even($+$) or odd ($-$) number of phonons, given electrons are a distance $\delta$ away (see Figure \ref{fig:parity-sketch}). The total wave function describing the system is a superposition of electron states with different relative distances of the two electrons. Depending on the relative distance, there is a preferred parity of the phonon number, defined by the expectation value \eqref{eqn:parity_operator}. If both the electron distance and the phonon number are even or they are both odd, the state belongs to the even parity sector $P_{+}$, otherwise, the state belongs to the odd parity sector $P_{-}$. We now consider the parity behavior in the anti-adiabatic ground state as the total momentum $\kappa$ is changed (see Figure \ref{fig:parity}). At $\kappa = 0$ (leftmost column) the ground state is a superposition of two states: (i) a state with an even phonon number with an even electron distance; (ii) a state with an odd phonon number with an odd electron distance. Therefore, as shown in Figure \ref{fig:parity-sketch}, the ground state at $\kappa = 0$ belongs to the even parity sector $P_+$. At an intermediate value of $\kappa = \pi/2$ (middle column), there is no definite parity, and at $\kappa = \pi$ (rightmost column) the ground state belongs to the $P_-$ parity sector. Applying the same arguments, it follows that the first excited state behaves opposite to the ground state, with parity changing from $P_-$ to $P_+$ as the total momentum is changed from $\kappa = 0$ to $\kappa = \pi$. Therefore, the first excited state can be understood as the opposite parity sector pair of the ground state. 

Given the above arguments, we see that the parity is not a property of the band. Instead, it is a property of a state in a band at a defined momentum $\kappa$. To understand the structure of the parity and the connection to the band structure, we consider the zero bare hopping limit $t=0$ of the Hamiltonian, where the parity symmetry is exact and the bands can be split into different parity sectors. In this case, the positive parity band has a monotonically increasing, while the negative parity band has a monotonically decreasing dispersion with momentum going from $\kappa = 0$ to $\kappa = \pi$ ~\cite{sous_light_2018}. The two bands cross at $\kappa = \pi/2$. Adding small bare hopping $t$ breaks the parity symmetry and strongly mixes the states around the crossing, while not affecting states away from it. As a result, the ground state changes the parity from $P_+$ to $P_-$ as the total momentum is swept, with undefined parity close to $\kappa = \pi/2$, while the first excited state does the opposite. The band splitting is also confirmed by the band dispersion, which exhibits an avoided crossing at $\pi/2$ (see Figure \ref{fig:excitation-spectrum}a). Interestingly, even though the hopping has a finite value, the parity symmetry is still significantly obeyed for $\kappa$ away from  $\pi/2$. This is a direct consequence of the change in electron momentum distribution between the weak and strong coupling regimes. At strong coupling the bare hopping gets significantly suppressed, giving rise to an effective parity symmetry. In the adiabatic regime, the parity is no longer a good quantum number, as the total bipolaron hopping is still dominated by the bare hopping (see Figure \ref{fig:parity}). Nevertheless, the first excited state is still present, with very similar binding properties.

\subsection{Higher bound states} \label{sec:higher-excited}

Apart from the parity complement of the ground state, there are also additional discrete bands at strong coupling (see Figure \ref{fig:excitation-spectrum}a-b). In the anti-adiabatic regime, both discrete bands have energies close to the lower-energy continuum edge, with states around $\kappa =0, \pi$ being below, and states around $\kappa = \pi/2$ being above the threshold. At all values of the total momentum, the electron-electron and electron-phonon correlation functions decay away from the bipolaron (see Figure \ref{fig:exc-correlations}a,c and Appendix \ref{sec:corr-finite-momentum} for more details), confirming these bands correspond to shallow bound states. Since the electron-electron correlation function does not change significantly, these states also correspond to the phonon excitations on top of the ground state bipolaron. Similarly to the first excited band, both discrete bands have a well-defined parity (see Figure \ref{fig:parity}). The second excited band has the same parity behavior as the first band, while the third excited band has the same parity as the ground state. In the adiabatic regime, the spectrum consists of four additional discrete bands (see Figure \ref{fig:excitation-spectrum}b). The second and the third excited bands correspond to strongly bound bipolaron-phonon states, while at zero total momentum, the highest two discrete bands correspond to shallow bound bipolaron-phonon states, as indicated by the correlation functions (see Figure \ref{fig:exc-correlations}b,d). At the intermediate value of the total momentum, $\kappa = \pi/2$, the highest discrete band is unbound, while the second highest band remains bound (see Appendix \ref{sec:corr-finite-momentum} for details), reflecting the weak binding energy of the highest excited states. The parity symmetry is broken, as for the first excited state (see Fig \ref{fig:parity}). 
In general, the excited states could have profound consequences on the dynamics of the bipolaron systems. Since they lie below the continuum, they are stable against single phonon decay and, unlike the ground state, have a much larger effective mass and a much smaller group velocity (see Figure \ref{fig:excitation-spectrum}a-b). For example, starting from an excited state, it will generally relax into a combination of the ground and excited states. The average propagation of such a system would then be much slower than what is predicted based on the ground state properties of the bipolaron. 

\subsection{Excitation continua}

In addition to the bound states analyzed previously, the spectrum also consists of two separate continua (see Figure \ref{fig:excitation-spectrum}a-b). The unbound nature of both continua can be seen in the electron-phonon correlations (see Figure \ref{fig:exc-correlations}a-b), which do not decay away from the bipolaron. The lower-energy continuum has excitation energies on the order of $\Omega$ and corresponds to a single-phonon continuum. This is confirmed by the correlation functions (see Figure \ref{fig:exc-correlations}a-d), which display significant change only in the electron-phonon correlation, while the electron-electron correlation remains the same. The continuum also contains a trivial zero momentum phonon excitation with energy exactly $E_{GS}(\kappa) + \Omega$.  In the SSH model the electrons couple only to the difference in phonon displacement, resulting in an electron-phonon coupling vanishing at zero phonon momentum, i.e. $g(k,q = 0) = 0$. Therefore, zero-momentum phonons, corresponding to a uniform displacement are decoupled, and the spectrum contains a trivial zero-momentum phonon excitation. This excitation is shown in the electron-phonon correlation function(see Figure \ref{fig:exc-correlations}a-b). Every site has the phonon number increased by $1/N$, the inverse system size. The higher-energy continuum lies close to the polaron continuum and corresponds to a mixed particle-hole and phonon continuum, as both correlation functions experience significant change (see Figure \ref{fig:exc-correlations}a-d). In the case of the bipolaron, the particle-hole excitations correspond to the deformation of the relative electron wavefunction. The exact nature of the mixing between the two excitation types depends on the selected state within the continuum, with higher energy states displaying a more significant change in the electron-electron correlations. Note, the \textit{ansatz} \eqref{eq:ansatz} with eight coherent states cannot capture the polaron continuum directly, as the unbound polaron state is not a linear excitation on top of the ground state. For a more general ansatz, the computed particle-hole-phonon continuum states will mix with the unbound polaron continuum. 

\subsection{Summary}

The main difference between the spectra in the adiabatic and the anti-adiabatic regime is the flat dispersion in the adiabatic regime, which is also reflected in the excitation spectrum. The discrete bands are flatter and the bandwidth of the two continua shrinks, collapsing the bands together (see Figure \ref{fig:excitation-spectrum}b). In Figure \ref{fig:excitation-spectrum}c we show the energy of the lowest four energy bands as a function of the coupling in the anti-adiabatic and adiabatic regimes, respectively. For both regimes, the excitations at low couplings are predominantly particle-hole excitations of the electrons, since the phonon energy is large. As the coupling is increased, the relevant energy scale for particle-hole excitations becomes $g^2/\Omega$ and the excitations become phonon-like. Furthermore, increasing the coupling also gives rise to bound excited states, which emerge below the two continua, starting from no bound states at low coupling to five/three bound states in the adiabatic/anti-adiabatic regime, respectively. Our results suggest that as the coupling, and hence the binding energy, is increased, new bound states emerge below the continuum. This is analogous to a particle in a square well, where increasing the potential depth gives rise to new bound states below the continuum. The computed excitation spectrum is qualitatively different compared to the Holstein model at strong coupling, where the lowest-lying excitation is a combination of the ground state bipolaron and a free phonon ~\cite{alexandrov_polaron_2000,devreese_frohlich_2015,thomas2024theoryelectronphononinteractionsextended}. Therefore, the rich structure of the low-energy excited states provides a signature of physics beyond the Holstein model~\cite{banerjee2024identifyingquantifyingsuschriefferheegerlikeinteractions}.

\section{Dynamical response of the bipolaron}\label{sec:response}

\begin{figure*}[t]
    \centering
    \includegraphics[width = 1\linewidth]{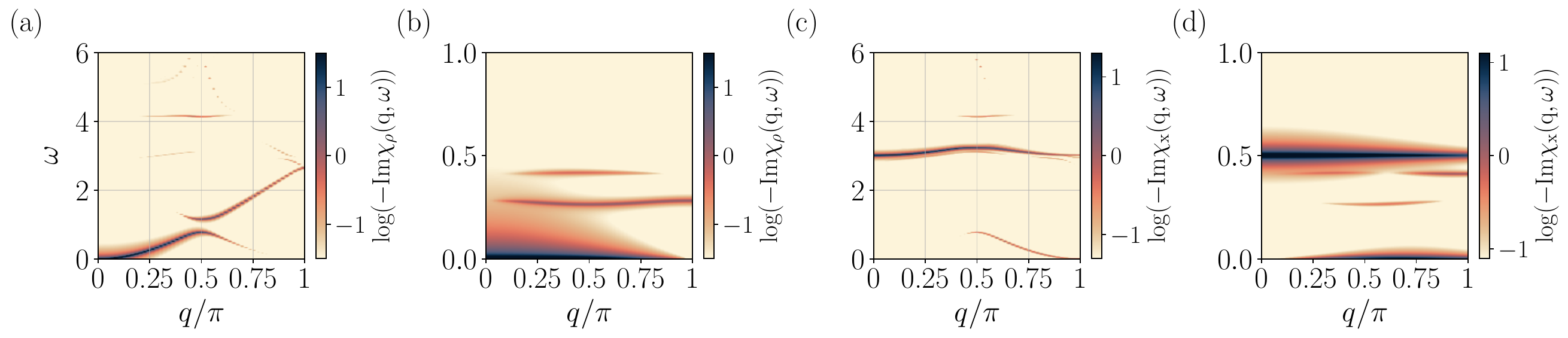}
    \caption{Dynamical density response function in units of inverse hopping as a function of external momentum $q$ and frequency $\omega$, for the bipolaron ground state at $\kappa = 0 $, computed using Lehman representation \eqref{eqn:response}, with artificial broadening of the delta function $\gamma = 0.01t$. a) Density response function \eqref{eq:density_response} for anti-adiabatic regime $\Omega = 3t$ at strong coupling $\lambda = 2.0$. b) Density response function \eqref{eq:density_response} for adiabatic regime $\Omega = 0.5t$ at strong coupling $\lambda = 0.8$. c) Displacement response function \eqref{eqn:displacement_response} for anti-adiabatic regime $\Omega = 3t$ at strong coupling $\lambda = 2.0$. d) Displacement response function \eqref{eqn:displacement_response} for adiabatic regime $\Omega = 0.5t$ at strong coupling $\lambda = 0.8$.}
    \label{fig:response} 
\end{figure*}

Having characterized the excitation spectrum and its properties, we now proceed to compute the density and the displacement linear response functions of the bipolaron. Using the formalism introduced before \eqref{eqn:response} and assuming the system is in the ground state, we use the Lehmann representation of the operator $\hat{\rho}_q =  \sum_{k,\sigma} \hat{c}^\dag_{k+q,\sigma} \hat{c}_{k,\sigma}$ to compute the response within the LLP frame

\begin{equation}
    \begin{split}
    \Im{\chi_{\rho}}&(q,\omega) = -\pi \sum_m \delta(\omega - \omega_{m,q})  \times \\
    &|\bra{m(q)} (1 + \sum_n \hat{c}^\dag_{n,\downarrow} \hat{c}_{n,\downarrow} e^{iqn} ) \ket{\rm{GS}(0)}|^2,
    \label{eq:density_response}
    \end{split}
\end{equation}
where $q,\omega$ are the momentum and the frequency of the linear response. The scalar term in the above equation $1$, independent of any operators, arises from the spin-up electron contribution in the LLP frame. Driving the density of the system at momentum $q$ couples different $\rm{LLP}$ sectors with different momenta, so the matrix element has to be evaluated between the ground state at zero momentum and the $m$-th eigenstate at total momentum $q$. Figure \ref{fig:response}a-b shows the density response function for $\lambda =2$, $\Omega = 3t$ and $\lambda = 0.8$, $\Omega = 0.5t$, respectively. The transitions are only allowed between certain bands. In the anti-adiabatic regime, for momentum $q < \pi/2$, the response is dominated by the transition to the ground state band, while for the $q > \pi/2$ it is dominated by the transition to the first excited band. Transitions to higher bands are suppressed, due to the large phonon numbers, which reduces the overlap of different states and the matrix element in equation \ref{eq:density_response}. To understand the overlap with the low-energy states, we consider the parity of the bands. The ground state changes the parity going from $P_+$ to $P_-$ for $\kappa$ from $0$ to $\pi$, while the first excited state does the opposite. Since the density operator does not change the parity, it can only couple the ground state at zero momentum to states with positive parity $P_+$. Therefore, the density response only has finite matrix elements with the ground state below $\pi/2$ and the first band above $\pi/2$. The parity symmetry of the SSH model induces an effective \textit{parity selection rule}, which determines the allowed transitions in the response function. The selection rules could also have interesting consequences for the decay processes between the bands. For example, the transition of the first excited band into the ground state can only happen by breaking the parity, or by transferring a large momentum to the bipolaron. In the adiabatic regime, parity is not conserved, so the matrix element between different bands is significant and the response function demonstrates no selection rules. 

In addition to computing the density response, we also compute the phonon displacement response of the system, given by a dimensionless operator $\hat{x}_q = \hat{b}_{-q} + \hat{b}_q^\dag$. By transforming the operator into the LLP frame and applying equation \eqref{eqn:response}, the displacement response becomes
\begin{equation}
    \begin{split}
    \Im & \{\chi_{x}\}(q,\omega) = -\pi \sum_m \delta(\omega - \omega_{m,q})  \times \\
    &|\bra{m(q)} (\hat{b}_{-q} - \hat{b}_q^\dag) \sum_n e^{iqn/2} \hat{c}_{n,\downarrow}^\dag \hat{c}_{n,\downarrow} \ket{\rm{GS}(0)}|^2,
    \label{eqn:displacement_response}
    \end{split}
\end{equation}
Figure \ref{fig:response}c-d shows the phonon displacement response function for $\lambda =2$, $\Omega = 3t$ and $\lambda = 0.8$, $\Omega = 0.5t$, respectively. The spectrum in the anti-adiabatic case can again be understood by considering the parity symmetry of the bands. The displacement operator creates/removes a phonon, changing the total parity of the system. Therefore, the ground state at zero momentum can only be connected to states with the opposite, i.e. negative parity $P_-$, and the displacement operator provides a complement to the density response spectrum. The displacement response also gets enhanced at higher momenta, reflecting the preferential coupling to the high momentum phonons, close to $\pm \pi$, and resulting in suppression of the matrix element close to $q = 0$ for all the bound states. As discussed before, the zero-momentum phonon excitation is an exact excitation of the Hamiltonian. Therefore, the matrix element of the displacement operator is very high, as shown by the response function. For the adiabatic regime, the parity is not conserved and we only observe the enhancement of matrix elements at large momentum values, along with the high overlap with the zero momentum excitation. 
The dimerization of the bands, combined with the parity selection rule is a smoking gun signature of the anti-adiabatic SSH-like bipolarons. For a single SSH polaron, the whole system is translationally invariant, so even and odd states are equivalent and the parity symmetry does not play an important role (see Appendix \ref{sec:figures}). The characteristic parity signature of bipolarons persists for couplings as low as $\lambda \approx 0.75$, which is approximately when the first excited state emerges below the continuum. Finally, the parity selection rule is also robust against adding additional phonon branches, as well as adding Coulomb interaction between the electrons. 

\section{Conclusion and outlook}
\label{sec:conclusion}
We have demonstrated that the momentum-dependent electron-phonon interaction, within the SSH model, leads to a more complex nature of the bipolaron states, compared to the canonical Holstein model ~\cite{alexandrov_polaron_2000,devreese_frohlich_2009}. At strong coupling, the individual electrons, which form the bipolaron, shift their momentum distribution to be around $\pm\pi/2$, consistent with the single polaron behavior~\cite{marchand_sharp_2010,shi_variational_2018,zhang_peierlssu-schrieffer-heeger_2021}. As a result, the bipolaron dispersion becomes almost doubly periodic, with minima at $0$ and $\pi$ total momentum. Furthermore, the momentum shift gives rise to utterly different effective mass behavior between the adiabatic and the anti-adiabatic regimes. Whereas in the anti-adiabatic regime, the bipolaron remains light~\cite{sous_light_2018}, in the adiabatic regime the effective mass increases exponentially. The non-local nature of the SSH coupling simply delays the exponential increase of the effective mass, which inevitably occurs at large phonon occupation numbers in all the regimes, as a consequence of the phonon cloud overlap.
The non-local nature of the SSH coupling also gives rise to an effective parity symmetry in the zero bare hopping limit. The Hilbert space gets split into two disconnected sectors; (i) superposition of odd electron distance with odd phonon number and even electron distance with even phonon number, and (ii) superposition of even electron distance with odd phonon number and odd electron distance with even phonon number. 
The excitation spectrum of the SSH model also demonstrates qualitatively different features compared to the Holstein model. In both adiabatic and anti-adiabatic regimes, there are multiple bound excited states below the single phonon continuum, which cannot decay to the lower energy states, resulting in infinitely lived bipolarons. As the coupling is increased, lower bound states emerge below the unbound polaron continuum, giving rise to a Feshbach resonance. Therefore, the bound states could have profound consequences on the nature of polaron interactions. In the anti-adiabatic regime, the first excited state is the opposite parity counterpart of the ground state and all the bound states have well-defined parity. As a result, the parity symmetry induces an effective \textit{parity selection rule} between different bands, forbidding certain transitions. This is directly observed in the density and the displacement response functions, which are constrained by the parity selection rule. Therefore, the response spectra provide a clear signature of the bipolaron formation in the SSH-like models. 

Our variational method can be generalized to the study of polaron and bipolaron dynamics, going beyond just ground-state physics. Given the constraint of the parity symmetry, as well as the stability of the bound states, the real-time dynamics of SSH bipolarons could provide further signatures of the bipolaron formation and give new insights into the dynamics of SSH-like systems~\cite{PhysRevLett.81.5382,di_formation_2011,mahajan_structure_2024}. The general form of the ansatz also allows applications to different types of electron-phonon systems, including systems with multiple phonon branches, or different types of coupling, enabling the modeling of polaron dynamics in certain cuprate or oxides systems~\cite{PhysRevB.102.235145}. 

Recently, there has been a lot of interest in studying exciton polaron systems~\cite{mahajan_structure_2024,dai_excitonic_2024}, where the interplay of exciton decay, Coulomb interaction, and polaron formation remains to be fully understood. All of these effects can be easily incorporated within our framework, allowing the study of real-time exciton-polaron dynamics and helping to understand the role of polarons in certain perovskite solar cell materials~\cite{troisi_charge-transport_2006,PhysRevLett.86.4624}. 

Finally, due to the efficiency of the ansatz, it can also be generalized to 2D systems~\cite{zhang_peierlssu-schrieffer-heeger_2021}, where different bipolaron pairing,  such as $d$-wave might emerge. Since our approach directly models the wavefunction, so different symmetry sectors can be studied separately, providing further understanding of the role momentum-dependent coupling plays in unconventional pairing systems, such as high-temperature superconductors. 

Apart from single bipolaron systems, in recent years there has also been interest in exploring the full many-body phase diagram ~\cite{zhang_bipolaronic_2023,tanjaroon_ly_comparative_2023,wang_robust_2022,xing_quantum_2021,weber_competing_2020,weber_excitation_2015,yirga_phonon-induced_2023,lim_route_2021,nocera_bipolaron_2021,fomichev_renormalized_2023}. With the full Holstein~\cite{lemmens_ground_1977,PhysRevLett.125.180602,PhysRevLett.66.778,PhysRevB.99.174516} and Holstein-Hubbard~\cite{PhysRevB.102.235145,PhysRevResearch.2.043258,nosarzewski_superconductivity_2021,PhysRevB.98.085405} phase diagrams only computed recently, the phase diagram of the SSH-like models remains an open question. Specifically, the question of the highest possible critical superconducting temperature and new routes toward superconductivity remains unanswered.  Our bipolaron results indicate a strong pairing of polarons with opposite spin, yielding a possible anti-ferromagnetic~\cite{yirga_phonon-induced_2023} or even a superconducting phase in the presence of coulomb interactions. More commonly, the SSH model is known to exhibit the charge-density-wave instability~\cite{xing_quantum_2021,weber_competing_2020, nocera_bipolaron_2021}. Therefore, the SSH electron-phonon model is expected to display a very rich phase diagram. The effect of the exact form of electron-phonon interaction and its momentum dependence on pairing is yet to be explored, leading to a large variety of possible phases~\cite{cai_antiferromagnetism_2021,kim_semiclassical_2024, gotz_phases_2024,yang_functional_2022,xing_attractive_2023,casebolt_magnetic_2024,cai_robustness_2022}. Understanding this effect could shed light on some of the puzzles in the high $T_c$ cuprate materials~\cite{tanjaroon_ly_comparative_2023,wang_robust_2022,lim_route_2021,zhang_bipolaronic_2023} and lead to new routes towards high-temperature superconductivity as well as provide new possible mechanisms of CDW formation~\cite{xing_quantum_2021, weber_competing_2020}.

\section*{Acknowledgements}
We acknowledge stimulating discussions with A. Gomez-Salvador, R. Andrei, I. Morera Navarro, I. Esterlis, T. Shi and thank J.B. Curtis for comments on the draft.  The ETH group acknowledges funding from the Swiss National Science Foundation project 200021\_212899, the Swiss State Secretariat for Education, Research and Innovation (contract number UeM019-1), and NCCR SPIN. E.D. acknowledges funding from ARO grant number W911NF-21-1-0184.

\onecolumngrid
\appendix
\newpage
\begin{center}
	\textbf{\Large Supplementary Materials}
\end{center}
\normalsize

\setcounter{equation}{0}
\setcounter{figure}{0}
\setcounter{table}{0}
\makeatletter
\setlength\tabcolsep{10pt}
\setcounter{secnumdepth}{2}

\section{LLP transformation and the derivation of the Hamiltonian}\label{sec:Hamiltonian}
\noindent In this section we summarize the effects of the two LLP transformations used in the main text 
\begin{equation}
    \hat{U}_{\rm LLP} = e^{-i\hat{X}_\uparrow(\hat{P}_\downarrow + \hat{Q}_b)}, \quad \hat{U}_{\rm LLP:2} = e^{i\pi \sum_q \hat{b}_q^\dag \hat{b}_q /2}e^{-i \hat{X}_\downarrow (\hat{Q}_b - \kappa)/2},
\end{equation}
where we define $\hat{X}_\sigma = \sum_n n \hat{c}_{n,\sigma}^\dag \hat{c}_{n,\sigma}$, $\hat{P}_\downarrow = \sum_k \hat{c}^\dag_{k,\downarrow} \hat{d}_{k,\downarrow}$, $\hat{Q}_b = \sum_q q \hat{b}^\dag_q \hat{b}_q$, $\kappa$ is the total momentum of the system and $\sigma$ denotes the spin of the electrons. Each operator is transformed according to $\hat{O} \rightarrow \hat{U}^\dag_{\rm LLP} \hat{O} \hat{U}_{\rm LLP }$, which for individual electron and phonon operators reduces to 
\begin{equation}
    \begin{split}
    \hat{c}_{n,\uparrow} &\rightarrow e^{-in(\hat{P}_\downarrow + \hat{Q}_b)}\hat{c}_{n,\uparrow} \\
\hat{c}_{k,\downarrow} &\rightarrow e^{-i\hat{X}_\uparrow k}\hat{c}_{k,\downarrow} \\ 
\hat{b}_q &\rightarrow e^{-i\hat{X}_\uparrow q} \hat{b}_q,
    \end{split}
\end{equation}
while the transformation rule for any higher-order operator can be derived by transforming the individual constituents. The LLP transformation of creation and annihilation operators results in the appearance of fermionic operator exponentials. To proceed, we project the operators onto a single-particle subspace in each spin sector. This simplifies the exponential operators according to $e^{in\hat{P}_\downarrow} = \sum_k e^{ikn} \hat{c}^\dag_{k,\downarrow} \hat{c}_{k,\downarrow} = \sum_m \hat{c}^\dag_{m-n,\downarrow} \hat{c}_{m,\downarrow}$, which can be proved by expanding the exponential and normal ordering the number operators. Here, we use $\hat{c}_{k,\sigma} = \frac{1}{\sqrt{N}}\sum_n \hat{c}_{n,\sigma} e^{-ikn}$ as the definition of the operators in the momentum space. The unitary transformation $\hat{U}_{\rm LLP}$ brings the total Hamiltonian into the form $\sum_\kappa \hat{H}_{\rm LLP, \uparrow}(\kappa) \hat{c}^\dag_{\kappa,\uparrow} \hat{c}_{\kappa,\uparrow}$, where
\begin{equation}
  \begin{split}
        \hat{H}_{\rm LLP, \uparrow}(\kappa) =  \sum_q \Omega_q \hat{b}^\dag _q \hat{b}_q &+ \sum_{\delta}t_{\delta}e^{-i\kappa\delta} e^{i(\hat{P}_\downarrow + \hat{Q}_b)\delta} 
         + \sum_{q,\delta} e^{-i(\kappa+q/2)\delta} e^{i(\hat{P}_\downarrow + \hat{Q}_b)\delta}\tilde{g}_{\delta}(q) \hat{b}_q + \rm{H.c.}  \\
        &+\sum_{k,\delta}t_{\delta}\hat{c}^\dag_{k,\downarrow} \hat{c}_{k,\downarrow} e^{-ik\delta}+ 
        \sum_{m,n,k,q,\delta} \hat{c}^\dag _{k+q,\downarrow} \hat{c}_{k,\downarrow} e^{-i\delta(k-q/2)} \tilde{g}_{\delta}(q) \hat{b}_q + \rm{H.c},
        \label{eqn:Hamiltonian-one}
    \end{split}
\end{equation}
 and we introduced the Fourier transform of the electron-phonon coupling in the LLP frame $\Tilde{g}_\delta(q) = \frac{2gi}{\sqrt{N}} \delta_{\delta,\pm1} \sin(\frac{q}{2})$. The transformation rule of the second unitary transformation $U_{\rm LLP:2}$ can similarly be written as
\begin{equation}
\begin{split}
    \hat{c}_{n,\downarrow} & \rightarrow \hat{c}_{n,\downarrow} e^{-in(\hat{Q}_b -\kappa)/2} \\
    \hat{b}_q & \rightarrow i \hat{b}_q e^{-i \hat{X}_\downarrow q/2}.
\end{split}    
\end{equation}
The final result of both transformations is to move the system to the center of mass frame. Due to finite lattice effects, it is not possible to do a single transformation, as the center of mass coordinate $R_{\rm CM} = \frac{r_1 + r_2}{2}$ is not guaranteed to belong to the lattice. The two-step unitary transformation allows us to exploit the total momentum conservation, using the first transformation, while respecting the symmetry between the two particles, using the second transformation. By applying the second unitary transformation $\hat{U}_{\rm{LLP}:2}$ and expanding the exponential operators in the single-fermion subspace we obtain the final Hamiltonian \eqref{eqn:Hamiltonian_final}. 

\section{Single polaron Hamiltonian}
 For a single polaron, we employ a similar transformation as the $\hat{U}_{\rm LLP}$, while removing any operators corresponding to the spin-down particle 
\begin{equation}
     \hat{U}_{\rm LLP:SP} = e^{-i\hat{X_\uparrow} \hat{Q}_b}.
\end{equation}

Applying the transformation to the Hamiltonian and expanding in the single particle sub-space gives, as derived in Ref ~\cite{shi_variational_2018}

\begin{equation}
    \hat{H}_{pol}(\kappa) = \Omega \sum_q \hat{b}^\dag_q \hat{b}_q + 2 t \cos(\kappa-\hat{Q}_b) + \frac{2ig}{\sqrt{N}} \sum_q \sin(\frac{q}{2}) \left( \cos(\kappa - \frac{q}{2} - \hat{Q}_b) \hat{b}_q - \cos(\kappa + \frac{q}{2} - \hat{Q}_b) \hat{b}^\dag_q  \right). 
    \label{eq:single-polaron-H}
\end{equation}

Note, that the above Hamiltonian only depends on the phonon operators, unlike in the bipolaron case, simplifying the analysis significantly. 

\section{Tangent space and Expectation values}\label{sec:expec}

 To construct the tangent space of the variational ansatz, we compute the partial derivatives of the state $\ket{\psi}$ w.r.t. the variational parameters $\alpha_{n}^c, \Delta_{c}(q)$ and $\Delta^*_{c}(q)$. These tangent vectors can be combined to compute the required Gram matrix $g_{\mu,\nu}$ and the projection of the Hamiltonian into the tangent space. The variational wave function has the form
 
\begin{equation}
    \ket{\psi} = \sum_c \sum_n \alpha_n \hat{c}_{n,\downarrow}^\dag \ket{0} \otimes e^{i \sum_q \left(\hat{b}^\dag_q \Delta_{c}(q) + \hat{b}_q \Delta_{c}^{*}(q) \right)} \ket{0}
    \label{eqn:ansatz_app}
\end{equation}

The tangent vectors then follow by taking derivatives w.r.t. individual parameters, keeping track of the commutation relations 

\begin{equation}
    \begin{split}
        \ket{\partial_{\alpha^c_n} \psi} &= \hat{c}^\dag_{n,\downarrow} \ket{0} \otimes \ket{\Delta^c} \\
          \ket{\partial_{\Delta_{c}(q)} \psi} &= \ket{\alpha_c} \otimes \left( \hat{b}^\dag_q - \frac{1}{2} \Delta^*_{c}(q) \right) \ket{\Delta_{c}}\\
         \ket{\partial_{\Delta_{c}^*(q)} \psi} &= \ket{\alpha_c} \otimes \left(-\frac{1}{2} \Delta_{c}(q)\right) \ket{\Delta_c}\\
    \end{split}
\end{equation}

where $\ket{\Delta_c} = e^{i \sum_q \left(\hat{b}^\dag_q \Delta_{c}(q) + \hat{b}_q \Delta_{c}^{*}(q) \right)} $ represents the $c$-th coherent state and we have acted with $\hat{b}_q$ on the coherent state, such that $\hat{b}_q \ket{\Delta_c} = \Delta_c(q) \ket{\Delta_c}$. Therefore, the tangent space consists of single-phonon excitations and particle-hole excitations of the electronic degree of freedom. We now demonstrate how to compute the expectation values of different operators for the \textit{ansatz} given by \eqref{eqn:ansatz_app}. We focus on computing the generic expectation value of the form 

\begin{equation}
    \bra{\Delta} \bra{\alpha} \hat{O} \ket{\bar{\alpha}} \ket{\bar{\Delta}},
\end{equation}

as any expectation value w.r.t. the full ansatz is obtained by summing over all the different electron and coherent state combinations. We introduce shortened vector notation $\hat{b}^\dag \Delta = \sum_q \hat{b}^\dag_q \Delta(q)$. Using this representation, we obtain the expectation values between two bosonic coherent states

\begin{equation}
    \begin{split}
        \langle  e^{i(\hat{b}^\dag \Bar{\Delta} + \hat{b} \Bar{\Delta} ^\dag )}e^{i(\hat{b}^\dag \Delta + \hat{b} \Delta ^\dag )}\rangle_0 &= e^{-\frac{1}{2}(\Bar{\Delta}_b^T \Bar{\Delta} + \Delta_b^T \Delta - 2 \Bar{\Delta}^\dag \Delta)} \\
        \langle e^{i(\hat{b}^\dag \Bar{\Delta} + \hat{b} \Bar{\Delta} ^\dag )} \hat{b}^\dag \hat{b} e^{i(\hat{b}^\dag \Delta + \hat{b} \Delta ^\dag )} \rangle_0 &=  \Bar{\Delta}^\dag \Delta \langle  e^{i(\hat{b}^\dag \Bar{\Delta} + \hat{b} \Bar{\Delta} ^\dag )}e^{i(\hat{b}^\dag \Delta + \hat{b} \Delta ^\dag )}\rangle_0\\
      \langle e^{i(\hat{b}^\dag \Bar{\Delta} + \hat{b} \Bar{\Delta} ^\dag )} e^{i \sum_q a_q \hat{n}_q} e^{i(\hat{b}^\dag \Delta + \hat{b} \Delta ^\dag )}\rangle_0 &= e^{-\frac{1}{2}(\Bar{\Delta}^T \Bar{\Delta} + \Delta^T \Delta - 2 \sum_q \Bar{\Delta}_{q}^\dag e^{ia_q}\Delta_{q}) } \\
      \langle e^{i(\hat{b}^\dag \Bar{\Delta} + \hat{b} \Bar{\Delta} ^\dag )} e^{i \sum_q a_q \hat{n}_q} \hat{b}_p  e^{i(\hat{b}^\dag \Delta + \hat{b} \Delta ^\dag ) } \rangle_0 &= \Delta(p) \langle e^{i(\hat{b}^\dag \Bar{\Delta} + \hat{b} \Bar{\Delta} ^\dag )} e^{i \sum_q a_q \hat{n}_q} e^{i(\hat{b}^\dag \Delta + \hat{b} \Delta ^\dag )}\rangle_0 \\
      \langle e^{i(\hat{b}^\dag \Bar{\Delta} + \hat{b} \Bar{\Delta} ^\dag )} e^{i \sum_q a_q \hat{n}_q} \hat{b}^\dag_p   e^{i(\hat{b}^\dag \Delta + \hat{b} \Delta ^\dag ) } \rangle_0 &= e^{i a_p }\Bar{\Delta }^*(p) \langle e^{i(\hat{b}^\dag \Bar{\Delta} + \hat{b} \Bar{\Delta} ^\dag )} e^{i \sum_q a_q \hat{n}_q} e^{i(\hat{b}^\dag \Delta + \hat{b} \Delta ^\dag )}\rangle_0\\
    \end{split}
\end{equation}

where $\expval{\cdots }_0 $ refers to expectation value w.r.t. phonon vacuum. To obtain an expectation value, we first normal order the operators, such that all the $\hat{b}^\dag$ act on the bra and $\hat{b}$ act on a ket, giving a c-number. The exponential factors of the form $\expval{e^{i\hat{n}_q a_q}}$ can be evaluated by noting that they act to rotate the coherent state $e^{i\hat{n}_q a_q} \ket{\Delta} = \ket{\Delta'}$, where $\Delta'(q) = e^{ia_q}\Delta_q$. The above terms provide all the expectation values that appear in the Hamiltonian and any observable of interest,  allowing for efficient and straightforward computations. The energy expectation value in the final frame \eqref{eqn:Hamiltonian_final} can be written as $ E = \sum_{c,c'} E_{c,\bar{c}}$, where $c,\bar{c}$ represent specific coherent states and $E_{c,\bar{c}}$ is 

\begin{equation}
\begin{split}
    E_{c,\bar{c}} = &\Omega \sum_q \bar{\Delta}^T \cdot \Delta e^{-\frac{1}{2}(\Bar{\Delta}_b^T \Bar{\Delta} + \Delta_b^T \Delta - 2 \Bar{\Delta}^\dag \Delta)} 
    \\ &-2t \sum_{n,\delta} \bar{\alpha}_{n-\delta}^* \alpha_{n} \left( e^{-\frac{1}{2}(\Bar{\Delta}^T \Bar{\Delta} + \Delta^T \Delta - 2 \sum_q \Bar{\Delta}^*(q) e^{i q/2}\Delta(q)) } e^{-i\kappa/2} + e^{-\frac{1}{2}(\Bar{\Delta}^T \Bar{\Delta} + \Delta^T \Delta - 2 \sum_q \Bar{\Delta}^*(q) e^{-i q/2}\Delta(q))}e^{i\kappa/2}  \right)\\
    & -\frac{4g}{\sqrt{N}} \sum_{q,n,\delta} \sin(q/2) \bar{\alpha}_{n-\delta}^* \alpha_{n} \times \\
    &\left[ \left( e^{-\frac{1}{2}(\Bar{\Delta}^T \Bar{\Delta} + \Delta^T \Delta - 2 \sum_q \Bar{\Delta}^*(q) e^{i q/2}\Delta(q)) } e^{-i\kappa/2} + e^{-\frac{1}{2}(\Bar{\Delta}^T \Bar{\Delta} + \Delta^T \Delta - 2 \sum_q \Bar{\Delta}^*(q) e^{-i q/2}\Delta(q))}e^{i\kappa/2}   \right) \right.\\
    &\cos\left( \frac{q}{2}(1-n\delta)\right) (\Delta(q)  + \bar{\Delta}^*(q))  \\
    &+ \left( e^{-\frac{1}{2}(\Bar{\Delta}^T \Bar{\Delta} + \Delta^T \Delta - 2 \sum_q \Bar{\Delta}^*(q) e^{i q/2}\Delta(q)) } e^{-i\kappa/2} - e^{-\frac{1}{2}(\Bar{\Delta}^T \Bar{\Delta} + \Delta^T \Delta - 2 \sum_q \Bar{\Delta}(q)^* e^{-i q/2}\Delta(q))}e^{i\kappa/2}  \right)\\
    &\left. \sin\left( \frac{q}{2} (1-n\delta)\right) (\bar{\Delta}^*(q) - \Delta(q)) \right]
\end{split}
\end{equation}
where we define a shorthand notation $\bar{\Delta}(q) = \Delta_{\bar{c}}(q)$, $\Delta(q) = \Delta_{c}(q)$, $\bar{\alpha}_n = \alpha_{\bar{c},n}$ and $\alpha_n = \alpha_{c,n}$, with $\bar{\Delta}$ and $\Delta$ representing the coherent state vector, as defined before. Similarly, the Gram matrix can also be evaluated in a block form, where each block corresponds to a different pair of coherent states $G_{c,\bar{c}}$, where 
\begin{equation}
    G_{c,\bar{c}} = \begin{pmatrix}
        G_{\bar{\alpha},\alpha} & G_{\bar{\alpha}, \Delta} & G_{\bar{\alpha},\Delta^*} \\
        G_{\bar{\Delta}, \alpha} & G_{\bar{\Delta}, \Delta} & G_{\bar{\Delta}, \Delta^*} \\
        G_{\bar{\Delta}^*, \alpha} & G_{\bar{\Delta}^*, \Delta} & G_{\bar{\Delta}^*, \Delta^*} \\
    \end{pmatrix}
\end{equation}

where the indices indicate the tangent vectors used in the overlap, which can be $\alpha_n$, $\Delta(q)$, or $\Delta^*(q)$. The individual components are then given as
\begin{eqnarray}
    G_{\bar{\alpha}_n,\alpha_{n'}} &=& \delta_{n,n'} e^{-\frac{1}{2}(\Bar{\Delta}_b^T \Bar{\Delta} + \Delta_b^T \Delta - 2 \Bar{\Delta}^\dag \Delta)} \\ 
    G_{\bar{\alpha}_n, \Delta(q')} &=& \bar{\alpha}_n \left(\bar{\Delta}^*(q') - \frac{1}{2}\Delta^*(q')\right) e^{-\frac{1}{2}(\Bar{\Delta}_b^T \Bar{\Delta} + \Delta_b^T \Delta - 2 \Bar{\Delta}^\dag \Delta)} \\
    G_{\bar{\Delta}(q), \Delta(q')} &=& \left(\sum_n \bar{\alpha}_n \alpha_{n}\right)\left[ \delta_{q,q'} +   \left(\Delta(q) - \frac{1}{2}\bar{\Delta}(q)) (\bar{\Delta}^*(q') - \frac{1}{2}\Delta^*(q')\right) \right] e^{-\frac{1}{2}(\Bar{\Delta}_b^T \Bar{\Delta} + \Delta_b^T \Delta - 2 \Bar{\Delta}^\dag \Delta)} \\ 
    G_{\bar{\alpha}_n, \Delta^*(q')} &=& -\bar{\alpha}_n \frac{1}{2}\Delta(q') e^{-\frac{1}{2}(\Bar{\Delta}_b^T \Bar{\Delta} + \Delta_b^T \Delta - 2 \Bar{\Delta}^\dag \Delta)} \\
    G_{\bar{\Delta}(q), \Delta^*(q')} &=& -\left(\sum_n \bar{\alpha}_n \alpha_{n}\right) \frac{1}{2} \Delta(q') \left(\Delta(q) - \frac{1}{2}\bar{\Delta}(q)\right) e^{-\frac{1}{2}(\Bar{\Delta}_b^T \Bar{\Delta} + \Delta_b^T \Delta - 2 \Bar{\Delta}^\dag \Delta)} \\ 
    G_{\bar{\Delta}^*(q), \Delta^*(q')} &=& \left(\sum_n \bar{\alpha}_n \alpha_{n}\right) \frac{1}{4} \Bar{\Delta}^*(q) \Delta(q') e^{-\frac{1}{2}(\Bar{\Delta}_b^T \Bar{\Delta} + \Delta_b^T \Delta - 2 \Bar{\Delta}^\dag \Delta)} \\ 
\end{eqnarray}

    where $\delta_{q,q'}$ represents the Kronecker delta function and the remaining entries in the Gram matrix can be evaluated by re-labeling $\alpha \rightarrow \bar{\alpha}$ and $\Delta \rightarrow \Bar{\Delta}$.

\section{Correlation functions at finite total momentum} \label{sec:corr-finite-momentum}

\begin{figure*}[h]
    \centering
    \includegraphics[width = 0.7\linewidth]{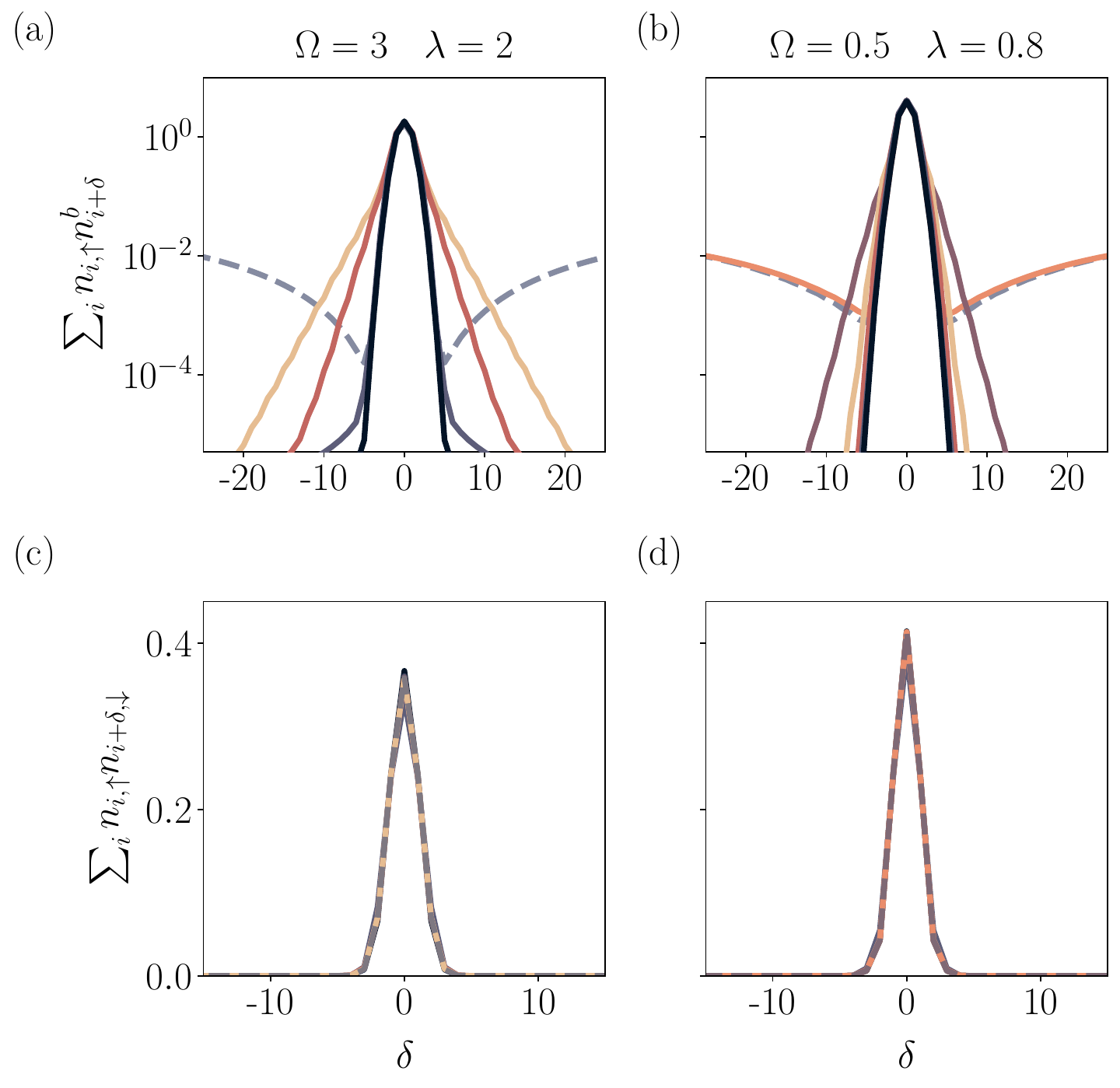}
    \caption{a) Electron-phonon correlation function $\expval{\sum_i n_{i,\uparrow} n^b_{i+\delta}}$ in the anti-adiabatic regime $\Omega =3t$ with $\lambda =2$ at total momentum $\kappa = \pi/2$, for the ground state (\textcolor{black1}{\sampleline{}}), 
        the three discrete excitations ((\textcolor{purple2}{\sampleline{}}),(\textcolor{red2}{\sampleline{}}),(\textcolor{yellow1}{\sampleline{}})) and a representative state of the single-phonon continuum (\textcolor{purple4}{\sampleline{dashed}}). 
        b)  Electron-phonon correlation function $\expval{\sum_i n_{i,\uparrow} n^b_{i+\delta}}$ in the adiabatic regime $\Omega =0.5t$ with $\lambda =0.8$ at total momentum $\kappa = \pi/2$, for the ground state (\textcolor{black1}{\sampleline{}}), the five discrete excitations ((\textcolor{purple2}{\sampleline{}}),(\textcolor{red2}{\sampleline{}}),(\textcolor{yellow1}{\sampleline{}}),(\textcolor{purple3}{\sampleline{}}),(\textcolor{red3}{\sampleline{}})) and a representative state of the single-phonon continuum (\textcolor{purple4}{\sampleline{dashed}}). 
         c) Electron-electron correlation function in the anti-adiabatic regime $\Omega =3t$ with $\lambda =2$ at  total momentum $\kappa = \pi/2$, for the ground state, discrete bands and a representative of the single-phonon continuum. Colors are the same as in a). d) Electron-electron correlation function in the adiabatic regime $\Omega =0.5t$ with $\lambda =0.8$ at total momentum $\kappa = \pi/2$, for the ground state, discrete bands and a representative of the single-phonon continuum. Colors are the same as in b).}
    \label{fig:correlation_app}
\end{figure*}

In this section we discuss the correlation functions at high coupling for the adiabatic and the anti-adiabatic regime at finite total momentum $\kappa = \pi/2$ (see Figure \ref{fig:correlation_app}). In both regimes, for $\kappa = 0$
all of the discrete excited states lie below the single-phonon excitation threshold, while for $\kappa = \pi/2$ the two highest discrete states lie above it. In the anti-adiabatic regime, the electron-phonon correlation functions decay for large separations, indicating that all the states are bound for all values of the total momentum. In the adiabatic regime, however, the electron-phonon correlation function of the highest excited state does not decay for large separations, indicating the state is unbound for $\kappa = \pi/2$. Therefore, for values of the total momentum close to $0,\pi$, the highest discrete excited state is bound, while for values of total momentum close to $\pi/2$ it is unbound. 

\section{Single polaron results}
\label{sec:figures}

\begin{figure*}[h]
    \centering
    \includegraphics[width = 0.7\linewidth]{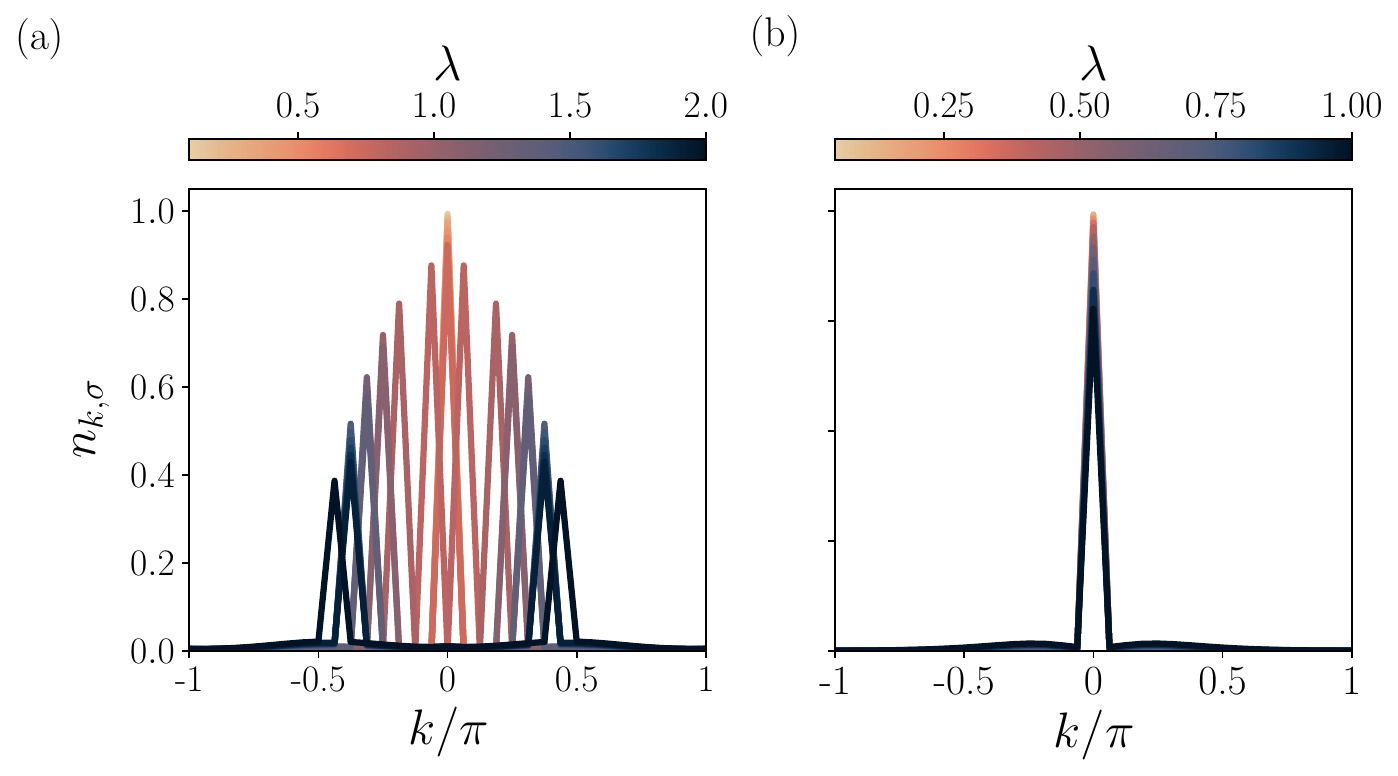}
    \caption{Single electron momentum distribution as a function of coupling $\lambda$, for the $\kappa = 0$ state. a) Momentum distribution for the anti-adiabatic regime $\Omega = 3t$ b) Momentum distribution for the adiabatic regime $\Omega = 0.5t$. }
    \label{fig:polaron_momentum}
\end{figure*}

In this section, we provide a summary of the single-polaron results for the SSH model. The results were computed using equivalent techniques as for the bipolaron. First, the single electron is integrated out using the $\rm LLP$ transformation, obtaining the LLP Hamiltonian as in Ref ~\cite{shi_variational_2018}. We then use a summed coherent state ansatz of the form
\begin{equation}
    \ket{\psi_{\rm sp}} = \sum_c^{N_c} a_c \ket{\Delta_c},
\end{equation}

where $a_c$ is the weight of the particular coherent state and $\Delta_c$ as defined before. The results shown are computed for $N = 32$ sites and $N_c = 20$ phonon coherent states. The single polaron momentum distribution can be seen in Figure \ref{fig:polaron_momentum}. At strong coupling, the electron momentum distribution is shifted towards $\pi/2$, as discussed in the main text, and can be seen in the anti-adiabatic case $\Omega = 3t$. For the range of couplings considered, the single polaron in the adiabatic regime $\Omega = 0.5t$ does not shift the momentum distribution towards the $\pi/2$. However, it was shown that at higher couplings the momentum shift still happens ~\cite{marchand_sharp_2010}. The critical coupling at which the phonon-assisted hopping becomes relevant is higher than for the bipolaron. This is expected, since in the bipolaron case both electrons can benefit from the phonon cloud, lowering the threshold.  

\begin{figure*}
    \centering
    \includegraphics[width = 0.7\linewidth]{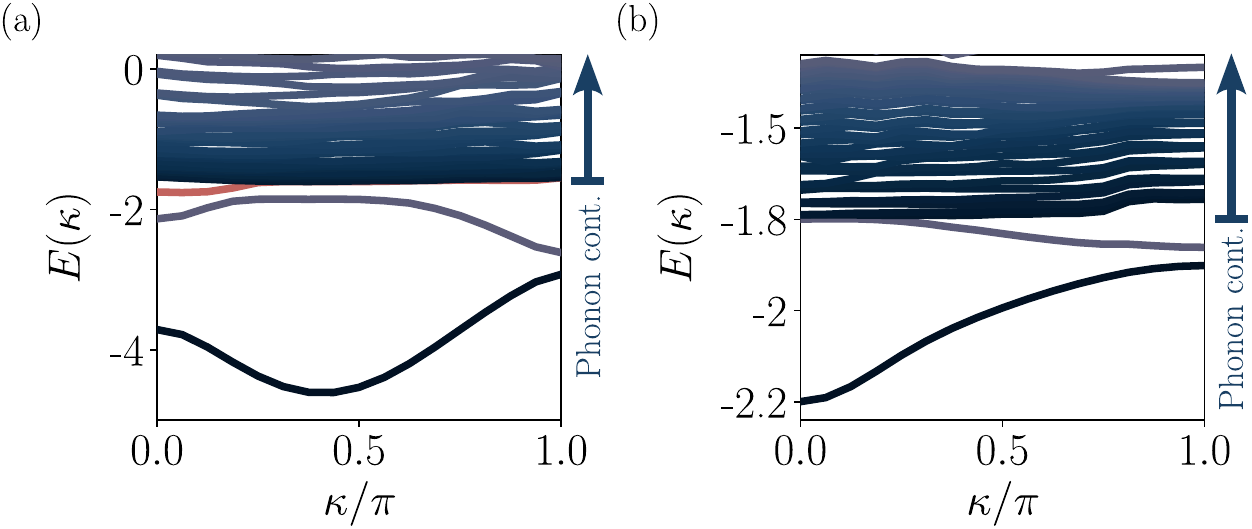}
    \caption{Excitation spectrum $E(\kappa)$ of the single polaron at strong coupling as a function of the total momentum, in units of hopping $t$. a) Spectrum for $\Omega = 3t$, $\lambda = 2$. b) Spectrum for $\Omega = 0.5t$, $\lambda = 0.8$.}
    \label{fig:polaron_spectrum}
\end{figure*}    

The spectrum of a single polaron displays a minimum at $ \kappa = \pi/2$, as reported previously ~\cite{marchand_sharp_2010,shi_variational_2018} (see Figure \ref{fig:polaron_spectrum}). Similar to the bipolaron case, there are excited states below the single phonon continuum in the strong coupling regime, corresponding to localized polaron-phonon states. Since the phonons are the only degree of freedom after the electron is integrated out, all excitations must be phononic in nature.
\begin{figure*}[h]
    \centering
    \includegraphics[width = 0.7\linewidth]{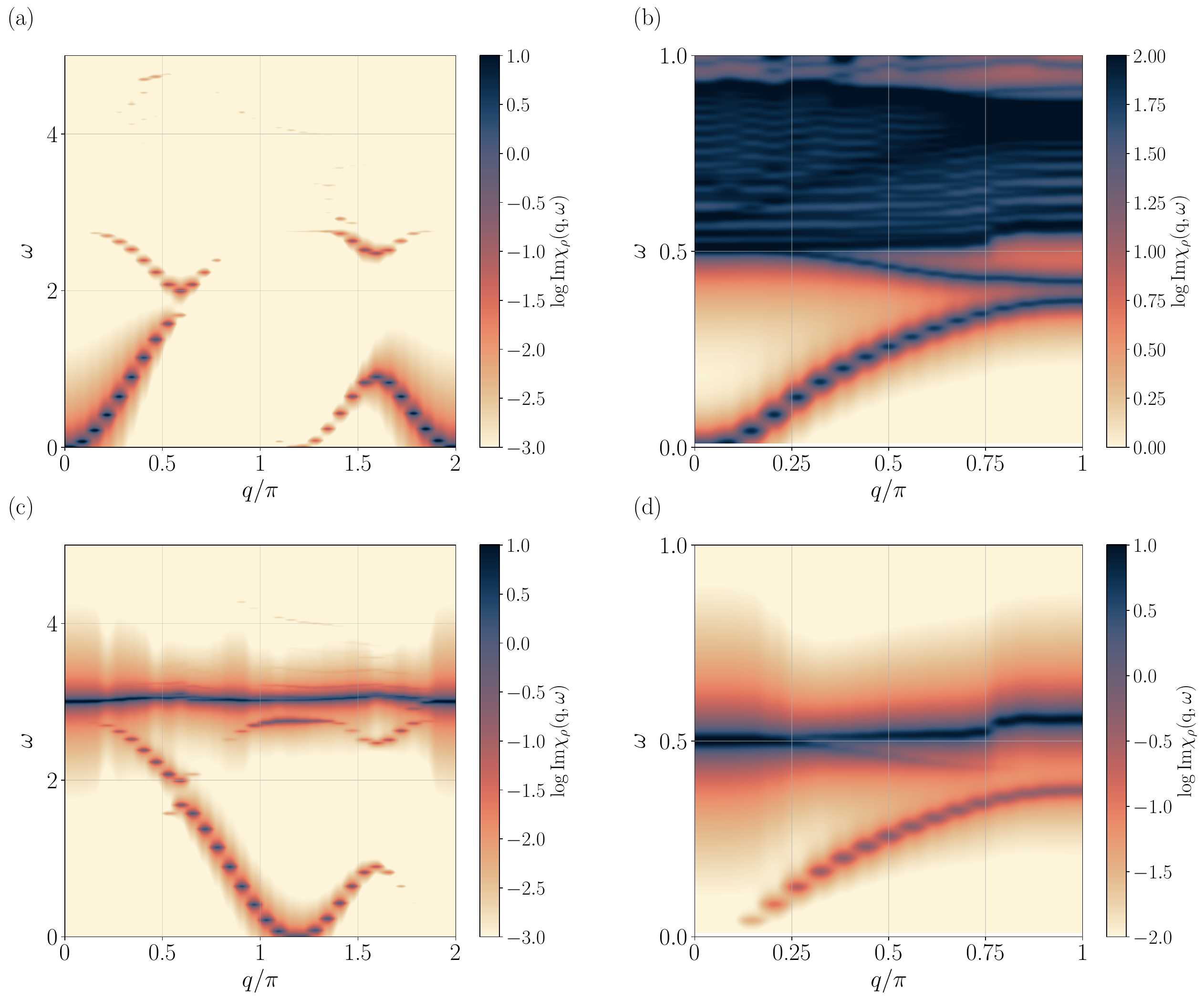}
    \caption{Response functions as a function of external momentum $q$ and frequency $\omega$, for the single polaron ground state, computed using Lehman representation \eqref{eqn:response}, with broadening of the delta function $\gamma = 0.01t$. a) Density response function \eqref{eq:density_response} for anti-adiabatic regime $\Omega = 3t$ at strong coupling $\lambda = 2$, starting from $\kappa = \pi/2$ ground state. b) Density response function \eqref{eq:density_response} for adiabatic regime $\Omega = 0.5t$ at strong coupling $\lambda = 0.8$, starting from $\kappa = 0 $ ground state. c) Displacement response function \eqref{eqn:displacement_response} for anti-adiabatic regime $\Omega = 3t$ at strong coupling $\lambda = 2.0$, starting from $\kappa = \pi/2$ ground state. d) Displacement response function \eqref{eqn:displacement_response} for adiabatic regime $\Omega = 0.5t$ at strong coupling $\lambda = 0.8$, starting from $\kappa = 0 $ ground state. }
    \label{fig:polaron_response}
\end{figure*}    
Having discussed the spectral properties, we now compute the response functions for the single polaron. In the anti-adiabatic regime, the ground state corresponds to a finite momentum value, and the density response function breaks the mirror symmetry. Therefore, we plot the density response for all the values $ 0 < q < 2\pi$ (see Figure \ref{fig:polaron_response} a). Within the LLP frame, the density response probes the overlap of the ground state at momentum $\kappa_{\rm GS}$ and states with momentum $\kappa_{\rm GS} + q$ 

\begin{equation}
    \Im{\chi_{\rho}}(q,\omega) = -\pi \sum_m \delta(\omega - \omega_{m,q}) \times |\bra{m(\kappa_{\rm GS} + q)}{\rm{GS}(\kappa_{\rm GS})}|^2.
    \label{eq:density_response-pol}
\end{equation}
The density-response functions show maxima for all states with a positive total value of momentum. Given the form of the single polaron Hamiltonian \eqref{eq:single-polaron-H}, we see that flipping the sign of the Hamiltonian and flipping all the phonon momenta $\hat{b}_q \rightarrow -\hat{b}_{-q}$ leaves the Hamiltonian invariant. 
Furthermore, for a given value of $\kappa = \pm\pi/2$, it can be shown that $\expval{\hat{b}_q} \approx \expval{\hat{b}_{-q}}$, as well as $ \expval{\hat{b}_q}= - \expval{\hat{b}^\dag_{q}} $. The former equality comes by considering the interaction term in the Hamiltonian, $\cos(\kappa - \hat{Q}_b - \frac{q}{2}) = \cos(\kappa - \hat{Q}_b) \cos(q/2) + \sin(\kappa - \hat{Q}_b)\sin(q/2)$ and ignoring $\cos(\kappa - \hat{Q}_b)$ term, while the latter comes as a consequence of the minus sign in \eqref{eq:single-polaron-H}. 
Therefore, the phonon occupations between $\kappa = \pi/2$ and $\kappa = -\pi/2$ are shifted from positive to negative momenta and flipped in sign, causing a small overlap between the states and suppressing the matrix element. Within the LLP frame, the displacement response is 

\begin{equation}
    \Im {\chi_{x}}(q,\omega) = -\pi \sum_m \delta(\omega - \omega_{m,q})  \times 
    |\bra{m(\kappa_{\rm GS} + q)} (\hat{b}_{-q} + \hat{b}_q^\dag) \ket{\rm{GS}(\kappa_{\rm GS})}|^2. 
    \label{eqn:displacement_response-pol}
\end{equation}

The maximum is obtained for the ground state of opposite total momentum.  As in the bipolaron case, the displacement response is enhanced for momenta close to $\pm \pi$, which corresponds to the most occupied phonon modes. Furthermore, since $\expval{\hat{b}_q} = - \expval{\hat{b}^\dag_q}$, the overlap is also maximized between $\kappa = \pi/2$ and $\kappa = - \pi/2$ ground states. In addition, since the zero momentum phonon is an exact excitation of the system, the overlap with the single phonon branch is very high, as in the bipolaron case. Overall, the polaron response function is asymmetric with the momentum $q$ and does not display the behavior of the parity selection rules, characteristic of the bipolaron. 

\end{document}